\documentclass[12pt]{article}
\usepackage[utf8]{inputenc}
\usepackage[english]{babel}
\usepackage{amsmath}
\numberwithin{equation}{section}
\usepackage{amsthm,amssymb}
\usepackage{bm}
\usepackage{amsfonts}
\usepackage[margin=1in,includefoot]{geometry}
\usepackage{fancyvrb}
\usepackage{dsfont}
\usepackage{mathtools}
\usepackage{mathrsfs}
\usepackage{float}
\usepackage{caption}
\usepackage{pgfgantt}
\usepackage{indentfirst}
\usepackage{tikz}
\usepackage[font=small,labelfont=bf]{caption}
\usepackage[nottoc]{tocbibind}
\usepackage[toc]{appendix}
\usepackage{algorithm}
\usepackage{multirow}
\usepackage{titlesec}
\usepackage[affil-it]{authblk}
\newcommand{\email}[1]{\texttt{\small #1}}
\allowdisplaybreaks

\newcommand{\E}{\mathbb{E}}

\newcommand{\1}{\mathbf{1}}

\newcommand{\V}{\mathbb{V}\text{ar}}

\newtheorem{proposition}{Proposition}

\author[]{Yiping Guo \footnote{Email: \email{y246guo@uwaterloo.ca} (Yiping Guo)}}
\author[]{Johnny Siu-Hang Li \footnote{Email: \email{shli@uwaterloo.ca} (Johnny Siu-Hang Li)}}
\affil[]{\normalsize Department of Statistics and Actuarial Science, University of Waterloo}

\usepackage[round]{natbib}
\usepackage{graphicx}
\graphicspath{{Pictures/}}
\usepackage{subfiles} 

\title{Robust Parameter Estimation for the Lee-Carter Family: A Probabilistic Principal Component Approach}
\date{}

\allowdisplaybreaks

\begin{document}
\maketitle
\begin{abstract}
	\noindent The well-known Lee-Carter model uses a bilinear form $\log(m_{x,t})=a_x+b_xk_t$ to represent the log mortality rate and has been widely researched and developed over the past thirty years. However, there has been little attention being paid to the robustness of the parameters against outliers, especially when estimating $b_x$. In response, we propose a robust estimation method for a wide family of Lee-Carter-type models, treating the problem as a Probabilistic Principal Component Analysis (PPCA) with multivariate $t$-distributions. An efficient Expectation-Maximization (EM) algorithm is also derived for implementation. 
 
    The benefits of the method are threefold: 1) it produces more robust estimates of both $b_x$ and $k_t$, 2) it can be naturally extended to a large family of Lee-Carter type models, including those for modelling multiple populations, and 3) it can be integrated with other existing time series models for $k_t$. Using numerical studies based on United States mortality data from the Human Mortality Database, we show the proposed model performs more robust compared to conventional methods in the presence of outliers.
\end{abstract}


\section{Introduction}
Understanding and modelling human mortality rates is essential for actuarial science and demography studies. Given the time series nature of the data, mortality rates are often studied using stochastic mortality models. Among many such models, the Lee-Carter model, first introduced by \cite{LC-1992} to forecast U.S. mortality rates, has become the ``leading statistical model of mortality forecasting in the demographic literature" according to \citep{Mor-2004-Deaton}. Initially designed to specifically project U.S. mortality rates, the method is now applied to various types of mortality data and is considered a benchmark in practice. The original Lee-Carter method consists of two main steps: parameter estimation and forecasting. The estimation step is based on singular value decomposition (SVD) to obtain the first principal component, and then a random walk model is used for forecasting.

Numerous extensions to the Lee-Carter model have been proposed over the years. For instance, \cite{EnhLC-2003-Renshaw} explored the feasibility of using multiple principal components to extend the deterministic estimation step. Another significant type of estimation technique is under the framework of generalized linear models (GLM), starting with the Poisson log-bilinear regression setup proposed by \cite{PoiLC-2002-Brouhns}. To tackle the potential overdispersion issue in the Poisson GLM, \cite{NBLC-2007-Delwarde} introduced a negative GLM structure. A comprehensive survey of the GLM framework of the Lee-Carter model can be found in \cite{GLMLC-2022-Azman}. Contrary to conventional Lee-Carter methods, which separate estimation and forecasting stages, \cite{BayesLC-2006-Pedroza} presented the Lee-Carter model as a Bayesian model, directly studying forecasting through the predictive posterior distribution. Recent literature has begun to leverage the power of modern machine learning to enhance model performance for multiple populations. \cite{DSA-2021-Diao} proposed a fully data-driven deletion-substitution-addition (DSA) algorithm to utilize information from multiple populations to predict mortality for a single population. Moreover, a neural network extension of the Lee–Carter model based on the paradigm of representation learning was developed by \cite{NNLC-2021-Richman} for forecasting mortality rates for multiple populations simultaneously.

While many extensions and improvements of the Lee-Carter method exist, its robustness in the presence of outliers has attracted limited attention. In actuarial science and statistics, the terms ``robustness" and ``outliers" can have multiple definitions. In the context of this study, we define ``outliers" as unusually high mortality rates that may arise due to rare but impactful events such as pandemics or wars. Such outliers are a natural occurrence in mortality data and can lead to unexpected shifts in long-term mortality trends. By ``robustness", we specifically refer to the resilience of parameter estimation against these outliers. In other words, a robust estimation method would ensure that the fitted parameters remain stable and reliable even when the mortality data contains such outliers. The traditional Lee-Carter model is designed primarily to model and forecast long-term mortality trends. Therefore, the effects of short-term outliers could distort these long-term projections and make them less reliable for actuarial and policy planning purposes.

In the literature, \cite{OutlierLC-2005-Li} conducted a systematic outlier detection and re-estimation process for the U.K. and Scandinavian countries, focusing on the time-series modelling for the mortality index $k_t$. A follow-up study on U.S. and Canadian mortality data was carried out by \cite{OutlierLC-2007-Li}. Additionally, \cite{HTLC-2011-Wang} suggested replacing the Gaussian innovations with several types of heavy-tailed errors in the Lee-Carter model to address potential outlier issues. Most existing literature on the outlier analysis of the Lee-Carter model focuses on the second stage: time series modelling for the mortality index $k_t$ while keeping the first parameter estimation stage unchanged. However, since the mortality indexes $k_t$ used to fit the time series model are directly obtained from the estimation stage, and the estimates of $a_x$ and $b_x$ are necessary for mortality forecasting, more attention should be given to robust parameter estimation in the first place.

This paper explores the robust parameter estimation of the Lee-Carter model by proposing a probabilistic principal component analysis (PPCA) model with multivariate $t$-distributions. The method is motivated by the fact that the traditional SVD approach used in the Lee-Carter estimation is equivalent to principal component analysis (PCA), which is well-known for its lack of robustness since the structure of the sample variance matrix can be profoundly influenced by extreme values. PPCA was first proposed by \cite{PPCA-1999-Tipping}, and it reformulates the conventional non-parametric PCA framework as a parametric Gaussian latent model, wherein the principal components can be exactly recovered by maximum likelihood estimation (MLE). Surprisingly, the PPCA formulation of the Lee-Carter model bears great resemblance to its state-space representation, which naturally encourages improving the robustness of the Lee-Carter estimation by robustifying the PPCA. The first robust PPCA model was proposed by \cite{RPPCA-2006-Archambeau}, replacing the Gaussian structure with multivariate $t$-distributions. Recently, \cite{MyPPCA-2022-Guo} introduced a more general formulation of multivariate $t$-distribution-based PPCA and the corresponding Monte-Carlo expectation-maximization algorithm.

To evaluate the performance of the proposed robust PPCA-based Lee-Carter model, we design real data experiments and simulations using data from \cite{HMD-2023-HMD}. The proposed method is first applied to U.S. mortality data across different time periods and compared with two traditional methods: SVD and Poisson GLM. We then create hypothetical outliers by adding the U.S. Covid-19 death numbers to the U.S. mortality data from 1970 to 2019. This design simulates scenarios of hypothetical pandemics occurring in history. The experimental results provide strong evidence that our estimation method performs significantly more robustly against outliers compared to the standard SVD and Poisson GLM approaches.

The remainder of the paper is organized as follows. Section 2 offers a review of the Lee-Carter model, standard PPCA, and their interconnection. Section 3 outlines the main method of this paper, including a quick introduction to the multivariate $t$-distributions, the proposal of the robust multivariate $t$-PPCA Lee-Carter model, an efficient expectation-maximization (EM) algorithm for its implementation, and parameter uncertainties. We discuss multi-population extensions in Section 4 to illustrate the versatility of our method. Section 5 then presents both real data illustrations and simulation studies. The conclusion and further remarks are given in Section 6.


\section{Background}
This section focuses on the Lee-Carter model and the standard PPCA, highlighting the intrinsic connection between these models, PCA, and PPCA. This discussion serves as the motivation for our proposed method, the robust multivariate $t$-PPCA, which is introduced in Section 3.

\subsection{The Lee-Carter Model}
The log central mortality rate for age group $x$ $(x=x_1,\cdots,x_p)$ at time $t$ $(t=t_1,\cdots,t_n)$ is denoted as $y_{x,t}:=\log(m_{x,t})$. The Lee-Carter model expresses $y_{x,t}$ in the following bilinear form:
\begin{equation}\label{Lee-Carter}
y_{x,t}:=\log(m_{x,t})=a_x+b_xk_t+\varepsilon_{x,t}.
\end{equation}
In this model, $a_x$ reflects the average level of the log mortality rate over time for age group $x$, while the time index $k_t$ is a time-specific parameter describing the overall mortality level at time $t$ for the studied population. Empirical evidence suggests the overall mortality rate is decreasing, hence $k_t$ usually follows a downward trend. The age-specific parameter $b_x$ measures the sensitivity of $y_{x,t}$ with respect to $k_t$. For example, a large $b_x$ implies the log mortality rate for age $x$ declines relatively slowly in response to the decline of $k_t$. $\varepsilon_{x,t}$ is a zero-mean error term, often assumed to be normally distributed. The constraints $\sum_x b_x=1$ and $\sum_t k_t=0$ are typically imposed to avoid identification issues.

Let $\hat{a}_x$, $\hat{b}_x$ and $\hat{k}_t$ be the estimates of the parameters $a_x$, $b_x$ and $k_t$. \cite{LC-1992} finds the least squares solution to \eqref{Lee-Carter}, and it immediately implies $\hat{a}_x=\Bar{y}_x:= \sum_t y_{x,t}/n$. They proposed to estimate $b_x$ and $k_t$ by SVD and the obtained solution can be interpreted as the first principal component of the log mortality rates. 

To gain more insights on the estimation, we adopt the following vector/matrix notations: $\bm{y}_{t}:=(y_{x_1,t},\cdots,y_{x_p,t})^T, \bar{\bm{y}}:=\sum_{t}\bm{y}_{t}/n, \bm{Y}:=(\bm{y}_{t_1},\cdots,\bm{y}_{t_n}), \bm{\Tilde{Y}}=(\bm{y}_{t_1}-\bar{\bm{y}},\cdots,\bm{y}_{t_n}-\bar{\bm{y}}), \bm{a}:=(a_{x_1},\cdots,a_{x_p})^T, \bm{b}:=(b_{x_1},\cdots,b_{x_p})^T, \bm{k}:=(k_{t_1},\cdots,k_{t_n})^T, \bm{\varepsilon}_t:=(\varepsilon_{x_1,t},\cdots,\varepsilon_{x_p,t})^T$, 
where $\bm{y}_{t},\bar{\bm{y}},\bm{a},\bm{b}$ and $\bm{\varepsilon}_t$ are $p$-dimensional column vectors, $\bm{k}$ is a $n$-dimensional column vector, while $\bm{Y}$ and $\bm{\Tilde{Y}}$ are $p\times n$ matrices, respectively. Thus, we can express \eqref{Lee-Carter} as:
\begin{equation}\label{Lee-Carter vector}
    \bm{y}_{t}=\bm{a}+\bm{b} k_t+\bm{\varepsilon}_t.
\end{equation}
As such, we can interpret the original least squares optimization as minimizing the squared reconstruction errors: 
\begin{equation}\label{Lee-Carter MSE}
    \min_{\bm{a},\bm{b},\bm{k}}\sum_{x,t}(y_{x,t}-(a_x+b_xk_t))^2= \min_{\bm{a},\bm{b},\bm{k}}\sum_{t}\Vert\bm{y}_t-(\bm{a}+\bm{b}k_t)\Vert_2^2,
\end{equation}
where $\Vert \cdot \Vert_2$ is the Euclidean norm (or equivalently, $L^2$ norm). 

Letting $\hat{\bm{a}}$ and $\hat{\bm{b}}$ be the estimates of the parameter vectors $\bm{a}$ and $\bm{b}$, standard PCA theory \citep{PRML-2006-Bishop} gives the optimal solution of the optimization problem \eqref{Lee-Carter MSE}:
\begin{equation}\label{Lee-Carter solution}
    \hat{\bm{a}}=\bar{\bm{y}}, \quad \Hat{\bm{b}}=\frac{\bm{u}}{\1^T\bm{u}},
\end{equation}
where $\1=(1,\cdots,1)$ is the $p$-dimensional column vector with all elements being 1, and $\bm{u}$ is the first left-singular vector of the centered log mortality data matrix $\Tilde{\bm{Y}}$ with unit length $\Vert \bm{u}\Vert_2=1$. The normalizing constant $\1^T\bm{u}$ is introduced to satisfy the identification constraint $\sum_x \hat{b}_x=1$, or equivalently $\1^T\Hat{\bm{b}}=1$. Then, the time index $k_t$ is found by matching the actual death numbers $D_{t}$ at time $t$. That is, for all $t=1,\cdots,n$, $\hat{k}_t$ is solution of the following equation
\begin{equation}\label{Lee-Carter k solution}
    D_t=\sum_x D_{x,t}=\sum_x \left(N_{x,t}\cdot e^{\hat{a}_x+\hat{b}_xk_t} \right),
\end{equation}
where $D_{x,t}$ and $N_{x,t}$ represent the total number of deaths and the risk of exposure of age group $x$ at time $t$, respectively. These equations can be solved numerically using standard root-finding methods such as one-dimensional line search. 

As it has been extensively studied, standard PCA is highly susceptible to outliers \citep{Robust-2004-Huber}, with even a single extreme point potentially significantly distorting the quality of the low-dimensional approximation. As such, it can be anticipated that the Lee-Carter estimates provided by SVD also lack robustness when outliers are present, a point that will be highlighted in our numerical analysis.

Another well-known type of estimating method for the Lee-Carter model is the Poisson bilinear GLM framework:
\begin{equation}\label{GLM}
    D_{x,t}\sim \text{Poisson}(N_{x,t}m_{x,t}),\ \text{with } \log(m_{x,t})=a_x+b_xk_t,
\end{equation}
This approach is likelihood-based, thus the parameter estimation is done through MLE, in particular, finding $\bm{a},\bm{b},\bm{k}$ to maximize the following log-likelihood function:
\begin{equation}\label{GLM MLE}
    \ell(\bm{a},\bm{b},\bm{k})=\sum_{x,t}\left(D_{x,t}(a_x+b_xk_t)-N_{x,t}e^{a_x+b_xk_t} \right)+\text{const}.
\end{equation}
The estimation is achieved by employing a modified iterative Newton-Raphson method, initially proposed by \cite{IRWS-1979-Goodman}. Similar to SVD, the MLE of the Poisson GLM is also sensitive to outliers, as noted in \cite{OutlierPoi-1989-Kunsch} and \cite{OutlierGLM-1992-Morgenthaler}, for example. Further numerical demonstrations will be provided in Section 5.

Once the estimates of $k_t$ are obtained (by either SVD, GLM, or other methods), a drifted random walk model is suggested by \cite{LC-1992} to model and forecast the time series $k_t $:
\begin{equation}\label{Random walk}
    k_t=k_{t-1}+\theta+e_t,
\end{equation}
where the innovation term is usually assumed to be normally distributed. This paper concentrates solely on the estimation stage.

\subsection{Standard PPCA and Connections to the Lee-Carter Model}
This subsection offers a brief introduction to Probabilistic Principal Component Analysis (PPCA) and establishes its relationship with both standard PCA and the Lee-Carter model.  For the sake of notational simplicity, we focus on the case involving just one principal component, aligning with the standard Lee-Carter model's formulation. For a broader treatment of PPCA with multiple principal components, refer to foundational texts like \cite{PPCA-1999-Tipping, PRML-2006-Bishop}.

As established in Section 2.1, estimating $\bm{a}$ and $\bm{b}$ of the Lee-Carter model is equivalent to solving a PCA, which can be achieved via SVD as outlined in \eqref{Lee-Carter solution}. Specifically, $\hat{\bm{a}}:=\bar{\bm{y}}$ is the average log mortality rates, and $\hat{\bm{b}}:=\bm{u}/\1^T\bm{u}$ is a normalized first left-singular vector of the centered log mortality data matrix $\Tilde{\bm{Y}}$. 

Interestingly, this SVD solution can also be expressed as a maximum likelihood estimate of a probabilistic latent variable model. Consider the following probabilistic model:
\begin{equation}\label{Lee-Carter PPCA}
    \bm{y}_t \sim \mathcal{N}(\bm{a},\bm{b}\bm{b}^T+\sigma^2\bm{I}),
\end{equation}
where $(\bm{a},\bm{b},\sigma^2)$ is the model parameters. \cite{PPCA-1999-Tipping} showed that the MLE of $(\bm{a},\bm{b},\sigma^2)$ in \eqref{Lee-Carter PPCA} have the following close-form solution:
\begin{equation}\label{Lee-Carter PPCA solution}
    \Hat{\bm{a}}=\Bar{\bm{y}},\quad \Hat{\bm{b}}=\bm{u}\sqrt{\lambda_1-\Hat{\sigma}^2}, \quad \Hat{\sigma}^2=\frac{1}{p-1}\sum_{i=2}^p\lambda_i,
\end{equation}
where $\bar{\bm{y}}$ is the average log mortality rates and $\bm{u}$ is the first eigenvector of the sample covariance matrix $\bm{S}:=\Tilde{\bm{Y}}^T\Tilde{\bm{Y}}/n$. Importantly, $\bm{u}$ is equivalent to the first left-singular vector of $\Tilde{\bm{Y}}$ up to a scaling factor, so $\hat{\bm{b}}$ derived from PPCA in \eqref{Lee-Carter PPCA solution} spans the same one-dimensional principal subspace as its counterpart in the standard Lee-Carter solution \eqref{Lee-Carter solution}. It implies that the PPCA solution \eqref{Lee-Carter PPCA solution} provides the identical parameter estimates of $\bm{a}$ and $\bm{b}$ as the standard Lee-Carter solution \eqref{Lee-Carter solution}, after normalizing $\hat{\bm{b}}$ under the identification constraint $\1^T\hat{\bm{b}}=1$.

This alternative approach, known as PPCA, was initially introduced by \cite{PPCA-1999-Tipping}. The adoption of PPCA offers a notable advantage: the probabilistic framework allows us to target specific statistical properties, such as robustness of parameters, which conventional PCA cannot provide. As we will explore in Section 3, a slight modification to standard PPCA can significantly improve the robustness of parameter estimates.

There exists an equivalent formulation of the PPCA model \eqref{Lee-Carter PPCA} via a latent variable structure:
\begin{equation}\label{Lee-Carter PPCA conditional}
    \bm{y}_t|z_t  \sim\mathcal{N}(\bm{a}+\bm{b}z_t,\sigma^2\bm{I}),\quad    z_t  \sim\mathcal{N}(0,1),
\end{equation}
where $z_t$ is an unobserved latent variable that follows a standard normal distribution. Marginalizing out the latent variable $z_t$ recovers the marginal distribution of $\bm{y}_t$ as described in \eqref{Lee-Carter PPCA}. The proof is straightforward and relies on applying the conditioning and marginalization properties of the multivariate normal distributions, as laid out in \eqref{Normal-Normal hierarchy} and \eqref{Normal-Normal marginal} in Appendix A. Readers are directed to Appendix A for additional details on the properties of multivariate normal distributions, which will be frequently referenced in Section 3 and Appendix B.

At first glance, the latent model structure in \eqref{Lee-Carter PPCA conditional} may seem redundant, as $z_t$ is neither an observed variable nor a model parameter. However, it turns out that the conditional structure serves a useful purpose: it lays the groundwork for an alternative approach to deriving the MLE using the EM algorithm. The EM algorithm becomes especially useful when directly solving the MLE is intractable, such as in the more complicated robust $t$-PPCA model developed in this paper, which will be discussed in Section 3.2.

It is worth noting that the PPCA model only estimates $\bm{a}$ and $\bm{b}$ for the Lee-Carter model. Thus, $k_t$ still needs to be estimated separately using the death number matching \eqref{Lee-Carter k solution}. Then, the estimated $k_t$'s are fitted using a chosen time series forecasting method, for instance, \eqref{Random walk}.

It is crucial to clarify the role of ``normality" when applying the PPCA method to the Lee-Carter model, as shown in \eqref{Lee-Carter PPCA}. Although it assumes that log mortality rates are normally distributed, the focus is not on whether the data strictly follow to a normal distribution. Our ultimate goal is not to fit a multivariate normal model to the data; rather, we aim to achieve the same PCA solution within a more flexible probabilistic framework that can accommodate further generalizations. In essence, we are not imposing distributional constraints on the mortality data. This concept is analogous to the case of linear regression, where the least squares estimates for the parameters do not rely on any specific data distribution. Yet, those estimates happen to coincide with the MLE when a normal distribution is assumed for the data.

Formulating the Lee-Carter model as PPCA \eqref{Lee-Carter PPCA} is attractive, since it provides exactly the same solution as the SVD method (the standard PCA), but the likelihood-based nature allows for a more flexible probabilistic formulation for various modelling purposes such as improvement model robustness. The analytical expressions in \eqref{Lee-Carter PPCA solution} also serve as good initial values of the numerical optimization in other modified models, as we will point out in Section 3.3.


\section{Robust Multivariate $t$-PPCA Lee-Carter Estimation}
Just like conventional PCA, PPCA is highly sensitive to outliers, as discussed in \cite{RPPCA-2006-Archambeau}. When atypical mortality data points exist, possibly due to events such as wars or pandemics, the estimated Lee-Carter parameters $\bm{b}$ may significantly deviate. This, in turn, can impact the subsequent time series modelling and forecasting step for $k_t$, as these values are derived from the estimated values of $\bm{a}$ and $\bm{b}$.

A computationally efficient method to enhance the robustness of PPCA involves changing the marginal Gaussian distribution to the multivariate $t$-distribution. For the multivariate $t$-PPCA models, resilience against outliers has been investigated through both theoretical and numerical studies, as seen in \cite{RPPCA-2006-Archambeau, RPPCA-2009-Chen, MyPPCA-2022-Guo}.

In the section, we adopt this framework and propose a ``$t$-PPCA" method for estimating $(\bm{a},\bm{b})$ for the Lee-Carter model with improved parameter robustness.

\subsection{Multivariate $t$-Distributions}
Multivariate $t$-distributions are widely used in a variety of robust statistical modelling problems. Specifically, maximum likelihood estimates from probabilistic models involving $t$-distributions often present stronger robustness against extreme observations \citep{Robust-1989-Lange}. 

Let $\bm{y}$ be a $p$-dimensional random vector following a multivaraite $t$-distribution $\bm{y}\sim t_{\nu}(\bm{\mu},\bm{\Sigma})$, where $\bm{\mu}$ is the $p$-dimensional mean vector, $\bm{\Sigma}$ is the $p \times p$ symmetric positive definite scale matrix, and $\nu>0$ is the degrees of freedom. Then, the probability density function of $\bm{y}$ is the following \citep{t-2006-Kibria}:
\begin{equation}\label{Multivariate t pdf}
    f(\bm{y})=
\frac{\Gamma[(\nu+p) / 2]}{\Gamma(\nu / 2) \nu^{p / 2} \pi^{p / 2}|\mathbf{\Sigma}|^{1 / 2}}\left[1+\frac{1}{\nu}(\bm{y}-\boldsymbol{\mu})^{T} \bm{\Sigma}^{-1}(\bm{y}-\boldsymbol{\mu})\right]^{-(\nu+p) / 2},
\end{equation}
where the $|\mathbf{\Sigma}|$ is the determinant of the scale matrix $\bm{\Sigma}$, and $\Gamma(\cdot)$ denotes the gamma function. When the degrees of freedom $\nu$ tends to infinity, it recovers the multivariate normal distribution $\mathcal{N}(\bm{\mu},\bm{\Sigma})$. Some useful properties of the multivariate $t$-distributions can be found in Appendix A.

\subsection{Model Formulation}
The core idea of using the $t$-PPCA method for estimating parameters in the Lee-Carter model is straightforward, as we will outline now. The standard normal PPCA formulation of the Lee-Carter model \eqref{Lee-Carter PPCA} provides the same estimates of $\bm{a}$ and $\bm{b}$ as the SVD method, but these results are sensitive to outliers. Therefore, adopting the multivariate $t$-distributions enables the model to naturally accommodate outliers. The $t$-PPCA method first modifies the normal assumption in the standard PPCA Lee-Carter model \eqref{Lee-Carter PPCA} to a multivariate $t$-distribution. That is, for $t=t_1,\cdots,t_n$, 
\begin{equation}\label{Lee-Carter t-PPCA}
    \bm{y}_t \sim t_{\nu}(\bm{a},\bm{b}\bm{b}^T+\sigma^2\bm{I}),
\end{equation}
Solving the MLE (with normalization $\bm{b}$ by $\1^T\bm{b}=1$) of \eqref{Lee-Carter t-PPCA} provides us with a robust parameter estimate for $\bm{a}$ and $\bm{b}$ in the Lee-Carter model. After the MLE of $\bm{a}$ and $\bm{b}$ are obtained, $k_t$ can be found by \eqref{Lee-Carter k solution}.

The role of the $t$-distribution in \eqref{Lee-Carter t-PPCA} warrants careful interpretation. Similar to the normality assumption in the standard PPCA model as discussed in Section 2.2, the log mortality data $\bm{y}_t$ needs not strictly follow a multivariate $t$-distribution. The key advantage of employing the $t$-distributions is that the resulting MLE of $(\bm{a},\bm{b})$ from \eqref{Lee-Carter t-PPCA} are expected to be more robust against outliers, yet still preserve essential information about the principal component. To illustrate this idea, consider the scenario of fitting a simple linear model to a dataset $(x_1,y_1),\cdots,(x_n,y_n)$ where the last point $(x_n,y_n)$ is an outlier. The fitted line (via either least square or normal MLE) will be largely influenced by the outlier $(x_n,y_n)$, leading to the lack of robustness. By contrast, adopting the $t$-distribution and computing the resulting MLE would make he fitted line more representative of the non-outlier points $(x_1,y_1),\cdots,(x_{n-1},y_{n-1})$. This exemplifies what we mean by ``parameter robustness".

To find the MLE of $\bm{a}$ and $\bm{b}$ in the proposed $t$-PPCA Lee-Carter model \eqref{Lee-Carter t-PPCA}, we write down the probability density function of \eqref{Lee-Carter t-PPCA} by substituting the corresponding parameters into \eqref{Multivariate t pdf}: 
\begin{equation}\label{t-PPCA Multivariate t pdf}
    f(\bm{y}_t)=
\frac{\Gamma[(\nu+p) / 2]}{\Gamma(\nu / 2) \nu^{p / 2} \pi^{p / 2}|\mathbf{\Sigma}|^{1 / 2}}\left[1+\frac{1}{\nu}(\bm{y}_t-\boldsymbol{a})^{T} (\bm{b}\bm{b}^T+\sigma^2\bm{I})^{-1}(\bm{y}_t-\boldsymbol{a})\right]^{-(\nu+p) / 2}.
\end{equation}
Directly maximizing the log-likelihood from \eqref{t-PPCA Multivariate t pdf} is computationally intractable. A common strategy to solve the MLE involving multivariate $t$-distributions is to adopt their scale mixture Gaussian representations \citep{tEM-1995-Liu} and then apply the EM algorithm, in which in each iteration the computation becomes tractable.

For the robust multivariate $t$-PPCA Lee-Carter model \eqref{Lee-Carter t-PPCA}, we propose the following equivalent hierarchical structure, which will be used to derive the EM algorithm in Section 3.3.
\begin{proposition}
    \begin{equation}\label{Lee-Carter t-PPCA mixture Gaussian representation}
    \bm{y} \sim t_{\nu}(\bm{a},\bm{b}\bm{b}^T+\sigma^2\bm{I}) \iff 
    \left\{\begin{aligned}
    &\bm{y}|z,u \sim  \mathcal{N} \left(\bm{a}+\bm{b}z,\frac{\sigma^2\bm{I}}{u}\right),\\
    &\ z|u \ \ \sim \mathcal{N} \left(0,\frac{1}{u}\right),\\
    &\ \ u \quad \,\sim \mathrm{Gamma}\left(\frac{\nu}{2},\frac{\nu}{2}\right),
    \end{aligned}\right.
\end{equation}
\end{proposition}

\begin{proof}
    The proof of Proposition 1 is by a straightforward application of the results presented in Appendix A.

    First, using \eqref{Normal-Normal hierarchy} and \eqref{Normal-Normal marginal}, we can show that $\bm{y}|z,u$ is multivariate normally distributed:
    \begin{equation}\label{Lee-Carter t-PPCA p(y|u)}
    \bm{y}|z,u \sim  \mathcal{N} \left(\bm{a}+\bm{b}z,\frac{\sigma^2\bm{I}}{u}\right), z|u \sim \mathcal{N} \left(0,\frac{1}{u}\right) \iff \bm{y}|u \sim \mathcal{N}\left(\bm{a},\frac{\bm{b}\bm{b}^T+\sigma^2\bm{I}}{u} \right).
    \end{equation}
    Then, using \eqref{Normal-Gamma hierarchy} and \eqref{Normal-Gamma marginal}, we can show that $\bm{y}$ is multivariate $t$-distributed, which elaborates \eqref{Lee-Carter t-PPCA mixture Gaussian representation}:
\begin{equation}\label{Lee-Carter t-PPCA p(y)}
    \bm{y}|u \sim \mathcal{N}\left(\bm{a},\frac{\bm{b}\bm{b}^T+\sigma^2\bm{I}}{u} \right), u \sim \mathrm{Gamma}\left(\frac{\nu}{2},\frac{\nu}{2}\right)\iff \bm{y} \sim t_{\nu}(\bm{a},\bm{b}\bm{b}^T+\sigma^2\bm{I}).
\end{equation}
\end{proof}

\subsection{The Expectation-Maximization (EM) Algorithm}
We next derive a computationally efficient EM algorithm to find the MLE of $\bm{a}$ and $\bm{b}$ of our proposed $t$-PPCA Lee-Carter model \eqref{Lee-Carter t-PPCA}, without directly handling the challenging density function \eqref{t-PPCA Multivariate t pdf}. In the following paragraphs, we demonstrate the implementing procedure and present the key updating formulas. The detailed derivations of \eqref{EM complete likelihood expectation}-\eqref{M-step nu} can be found in Appendix B.

The EM algorithm was first introduced by \cite{EM-1977-Dempster} and has become a powerful iterative optimization technique to find the MLE involving missing data or latent variables. A general introduction to the EM algorithm can be found in many standard textbooks on machine learning, such as \cite{ESLII-2009-Hastie} and \cite{PRML-2006-Bishop}. Instead of maximizing the original log-likelihood function derived from \eqref{t-PPCA Multivariate t pdf}, the EM algorithm utilizes the hierarchical structure \eqref{Lee-Carter t-PPCA mixture Gaussian representation}, in which each conditional distribution is easier to deal with analytically. 

The first step is to write down the complete log-likelihood function for the observed log mortality rates $\bm{y}_{t}$, by assuming that the latent variables $z_t$ and $u_t$ are observed:
\begin{equation}\label{EM complete likelihood}
     L_c = \sum_{t} \log [p(\bm{y}_t, z_t, u_t)]=\sum_{t} \log[p(\bm{y}_t|z_t,u_t)p(z_t|u_t)p(u_t)],
\end{equation}
where $p(\bm{y}_t|z_t,u_t)$, $p(z_t|u_t)$ and $p(u_t)$ are the probability density functions of the corresponding distributions established in \eqref{Lee-Carter t-PPCA mixture Gaussian representation}.

In the E-step, we need to find the conditional expectation of the complete log-likelihood $\langle L_c \rangle$ conditioning on the observed data $\bm{y}_t$. Substituting the expressions of $p(\bm{y}_t|z_t, u_t)$, $p(z_t|u_t)$, and $p(u_t)$ into \eqref{EM complete likelihood}, we obtain:
\begin{align}\label{EM complete likelihood expectation}
    \langle L_c \rangle =& -\sum_{t} \bigg [\frac{p}{2}\log \sigma^2 +\frac{\langle u_{t}\rangle}{2\sigma^2}(\bm{y}_t-\bm{a})^T(\bm{y}_t-\bm{a})-\frac{1}{\sigma^2}\langle u_{t}z_t\rangle \bm{b}^T(\bm{y}_t-\bm{a})
    \nonumber \\
    &+\frac{1}{2\sigma^2}\bm{b}^T\bm{b}\langle u_{t}z_t^2\rangle-\frac{\nu}{2} \left(\log \frac{\nu}{2}+\langle \log u_{t}\rangle-\langle u_{t}\rangle \right)+\log \Gamma \left(\frac{\nu}{2} \right)\bigg]+\mathrm{const}.,
\end{align}
where $\langle \cdot \rangle=\E[\cdot|\bm{y}_t]$ denotes the conditional expectation operator. It turns out that all the posterior expectations in \eqref{EM complete likelihood expectation} have analytical forms, which makes the E-step computationally efficient:
\begin{align}
    \langle u_{t}\rangle&=\frac{\nu+p}{\nu+(\bm{y}_t-\bm{a})^T(\bm{b}\bm{b}^T+\sigma^2\bm{I})^{-1}(\bm{y}_t-\bm{a})},\label{E-step u} \\
    \langle \log u_{t}\rangle&=\psi \left(\frac{\nu+p}{2} \right)-\log \left(\frac{\nu+(\bm{y}_t-\bm{a})^T(\bm{b}\bm{b}^T+\sigma^2\bm{I})^{-1}(\bm{y}_t-\bm{a})}{2} \right),\label{E-step log u} \\
    \langle z_{t}\rangle&= (\bm{b}^T\bm{b}+\sigma^2)^{-1}\bm{b}^T(\bm{y}_t-\bm{a}), \label{E-step z}\\
    \langle u_tz_{t}\rangle&=\langle u_t\rangle \langle z_{t}\rangle, \label{E-step uz} \\
    \langle u_tz_{t}^2\rangle&=\sigma^2(\bm{b}^T\bm{b}+\sigma^2)^{-1}+\langle u_{t}\rangle \langle z_{t}\rangle^2, \label{E-step uz2}
\end{align}
where $\psi(\cdot)=\Gamma(\cdot)/\Gamma^{\prime}(\cdot)$ is the digamma function.

In the M-step, we maximize $\langle L_c \rangle$ with respect to the parameters $(\bm{a},\bm{b},\sigma^2,\nu)$, by setting all the first order partial derivatives to 0. This results in the following updating equations:
\begin{align}
    &\bm{a}\longleftarrow \frac{\sum_{t}\langle u_{t}\rangle \left( \bm{y}_t-\bm{b}\langle z_t\rangle \right) }{\sum_{t}\langle u_{t} \rangle},\label{M-step a} \\
    &\bm{b} \longleftarrow \left[\sum_{t}\langle u_{t}z_t^2 \rangle \right]^{-1}\left[\sum_{t}(\bm{y}_t-\bm{a})\langle u_{t}z_t \rangle \right],\label{M-step b} \\
    &\sigma^2\longleftarrow \frac{1}{np}\sum_{t}\Big[\langle u_{t}\rangle (\bm{y}_t-\bm{a})^T (\bm{y}_t-\bm{a})-2\langle u_{t}z_t \rangle \bm{b}^T (\bm{y}_t-\bm{a})+\bm{b}^T\bm{b}\langle u_{t}z_t^2\rangle \Big]\label{M-step sigma2}, \\
    &1+\log \frac{\nu}{2}-\psi \left(\frac{\nu}{2}\right)+\frac{1}{n}\sum_{t}\left(\langle \log u_{t} \rangle -\langle u_{t} \rangle \right)=0 \label{M-step nu},
\end{align}
where the updated solution of $\nu$ in \eqref{M-step nu}  can be found by using a one-dimensional line search.

The two-stage EM algorithm for finding the MLE of $(\bm{a},\bm{b})$ is implemented by alternating the E-step and M-step until convergence. Convergence is defined as the point at which the absolute difference in the log-likelihood between two successive iterations is less than a small threshold, which is chosen as $10^{-4}$ throughout this paper. It is worth noting that we need normalize $\hat{\bm{b}}$ by 
\begin{equation}\label{b normalization}
    \hat{\bm{b}}:= \frac{\hat{\bm{b}}}{\1^T\hat{\bm{b}}},
\end{equation}
to meet the identification constraint $\sum_x\hat{b}_x=1 $. Then,  $k_t$ can be found by \eqref{Lee-Carter k solution}. The primary steps for estimating the parameters in the robust multivariate $t$-PPCA Lee-Carter model are summarized in Algorithm 1.

The updating of $\bm{a}$ and $\bm{b}$ provide insightful interpretations as to why the resulting estimates are more robust. The update equations \eqref{M-step a} and \eqref{M-step b} can be understood as a weighted average and weighted least square solution, respectively, where $u_t$ serves as the weight parameter for observation $\bm{y}_t$. In \eqref{E-step u}, we have shown that
\begin{equation}\label{E[u|y]}
    \E[u_t|\bm{y}_t]=\langle u_{t}\rangle=\frac{\nu+p}{\nu+(\bm{y}_t-\bm{a})^T(\bm{b}\bm{b}^T+\sigma^2\bm{I})^{-1}(\bm{y}_t-\bm{a})},
\end{equation}
where the derivation can be found in Appendix B. When an observation $\bm{y}_t$ is a potential outlier, it should be located far away from the center $\bm{a}$, which results in a large ``distance" $(\bm{y}_t-\bm{a})^T(\bm{b}\bm{b}^T+\sigma^2\bm{I})^{-1}(\bm{y}_t-\bm{a})$. It implies that $u_t$ is more likely to be small and consequentially, \eqref{M-step a} and \eqref{M-step b} will down-weight the outlier during the updating of $\bm{a}$ and $\bm{b}$, respectively.

\begin{algorithm}[t]
\vspace{0.2cm}
\begin{enumerate}
    \item Initialize $\bm{a},\bm{b},\sigma^2$ and $\nu$. 
    
    \item Estimate $\bm{a}$ and $\bm{b}$ by the EM algorithm described in Section 3.2:
    \begin{enumerate}
        \item E-step: Given the current estimates of the parameters $(\bm{a},\bm{b},\sigma^2,\nu)$, compute the posterior conditional expectations \eqref{E-step u}-\eqref{E-step uz2} in the E-step.
        \item M-step: Update the parameters $(\bm{a},\bm{b},\sigma^2,\nu)$ by \eqref{M-step a}-\eqref{M-step nu} in the M-step.
        \item Repeat Step 2(a) and Step 2(b) until convergence. 
        \item Normalize $\hat{\bm{b}}$ by \eqref{b normalization} to satisfy the identification constraint $\sum_x\hat{b}_x=1 $.
    \end{enumerate}
    \item Estimate $k_t$ by death number matching \eqref{Lee-Carter k solution}
    
\end{enumerate}
\caption{Robust Multivariate $t$-PPCA Lee-Carter Estimation}
\end{algorithm}

\subsection{Parameter Uncertainty}
Understanding parameter uncertainties, specifically the standard errors of parameter estimates, is of critical importance in stochastic mortality models. In this regard, this subsection describes the methodology for assessing parameter uncertainties in our proposed $t$-PPCA model.

The classical Lee-Carter model \cite{LC-1992} recommended using a residual bootstrap procedure to approximate the variance of the parameter estimates. While analytical expressions are available for the standard errors derived from SVD, these are primarily relevant in scenarios where errors are attributable to sampling or measurement errors. However, in stochastic mortality modelling, errors frequently emerge due to the discrepancies between observed mortality patterns and the simplified model structure, such as the bilinear framework of the Lee-Carter model. This distinction makes the bootstrap approach particularly well-suited for capturing various sources of uncertainty, thereby providing a more robust estimate of parameter variance.

For the $t$-PPCA method proposed in this study, no closed-form analytical solutions exist for calculating the standard errors of the parameter estimates. However, given that the prediction errors for the log mortality rate $\bm{y}_t$ readily available after fitting the model, we adopt a similar nonparametric residual bootstrap strategy following \cite{LC-1992}. During each bootstrap iteration, a synthetic data matrix is generated. This is done by adding residuals to the fitted data matrix $\Hat{\bm{Y}}=(\Hat{\bm{y}}_1,\cdots,\Hat{\bm{y}}_n)$, where $\Hat{\bm{y}}_t=\hat{\bm{a}}+\hat{\bm{b}}\hat{k}_t$ is the fitted log mortality rate vector for all ages at time $t$. These residuals are obtained through a sampling process with replacement from the vectors of fitting errors $\bm{y}_{t}-\bm{\Hat{y}}_{t}$ across all $t$.

Subsequently, the parameter $\bm{a}$, $\bm{b}$, and $k_t$ are re-estimated using this synthesized data matrix. It is noteworthy that the synthetic data matrices generated in each bootstrap iteration serve as perturbed instances of the original dataset, thereby allowing for the quantification of parameter uncertainty. Finally, the standard errors for each parameter can be approximated by calculating the sample standard deviation from the bootstrap-derived parameter estimates.

\section{Multi-Population Extensions}
A significant feature of the proposed approach is its high generality, meaning it can be applied to a broad range of Lee-Carter type models. Whenever SVD is employed for parameter estimation, our approach can serve as a replacement for enhancing robustness. In this section, we briefly discuss the application of our method to two multi-population extensions of the Lee-Carter model.

\subsection{Augmented Common Factor (ACF) Model}
We first study the ACF model, proposed by \cite{multiLC-2005-Li} to forecast mortality rates for multiple populations. The produced mortality forecasts are called ``coherent", that is, do not diverge in the long term between different populations.  

Indexing the populations by $i=1,\cdots,I$, the ACF model decomposes the log mortality rates of population $i$ to a common factor and a specific factor:
\begin{equation}\label{ACF}
    y_{x,t,i}:=\log(m_{x,t,i})=a_{x,i}+\underbrace{b_xk_t}_{\text{common factor}} +\underbrace{b_{x,i}k_{t,i}}_{\text{specific factor}} ,
\end{equation}
or in the vector form:
\begin{equation}\label{ACF vector}
  \bm{y}_{t,i}:=\log(\bm{m}_{t,i})=\bm{a}_i+\bm{b}k_t+\bm{b}_ik_{t,i},  
\end{equation}
Similar to the original Lee-Carter model, the procedures involved in this method are roughly divided into two stages: 1) estimating the parameters $\bm{a}_i$, $\bm{b} $, $\bm{b}_i$, $k_t$ and $k_{i,t}$, and then 2): building time series models and making forecasting for $k_t$ and $k_{t,i}$. 

Here we focus on the estimation stage since it is where our $t$-PPCA approach can be applied. The conventional approach for the estimation can be done through three steps:
\begin{enumerate}
    \item For each of $i=1,\cdots,I$, take 
    \begin{equation}\label{ACF a}
        \hat{\bm{a}}_{i}=\Bar{\bm{y}}_{i}:=\frac{1}{n}\sum_t \bm{y}_{t,i},
    \end{equation}
    the average log mortality rates for population $i$. Also compute the centered log mortality rates at time $t$,  $\Tilde{\bm{y}}_{t,i}=\bm{y}_{t,i}-\Bar{\bm{y}}_{i}$.
    \item Estimate the common factors $\bm{b}$ and $k_t$ from the average centered log death rates by SVD:
    \begin{equation}\label{ACF common b}
        \Tilde{\bm{y}}_t =\bm{b}k_t,
    \end{equation}
    where $\Tilde{\bm{y}}_t:=\sum w_i \Tilde{\bm{y}}_{t,i}$ is computed as a weighted average, and the weights $w_1,\cdots,w_I$ are determined by the population sizes as in \cite{multiLC-2005-Li}.  
    \item For population $i=1,\cdots,I$, estimate the specific factors $\bm{b}_i$ and $k_{t,i}$ from the residuals by SVD:
    \begin{equation}\label{ACF individual b and k}
        \underbrace{\bm{y}_{t,i}-\hat{\bm{a}}_i}_{\Tilde{\bm{y}}_{t,i}} -\hat{\bm{b}}\hat{k}_t=\bm{b}_ik_{t,i},
    \end{equation}
    where $\hat{\bm{b}}$ and $\hat{k}_t$ are the estimates obtained in Step 2. 
\end{enumerate}

Both Steps 2 and 3 use standard SVD for estimation, thus we can simply replace it with the $t$-PPCA approach. It is worth noting that our proposed $t$-PPCA method is designed to fit the original log mortality rates, via the framework $\bm{y}\sim t_{\nu}(\bm{a},\bm{b}\bm{b}^T+\sigma^2\bm{I})$, where the mean parameter $\bm{a}$ needs to be estimated as a parameter. However, we do not need to estimate this parameter in both \eqref{ACF common b} and \eqref{ACF individual b and k}, as $\hat{\bm{a}}_i$ has been calculated by \eqref{ACF a} in Step 1. This implies that we are in essence fitting a restricted $t$-PPCA model $\bm{y}\sim t_{\nu}(\bm{0},\bm{b}\bm{b}^T+\sigma^2\bm{I})$ to the ``residuals", with the mean parameter $\bm{a}$ fixed as $\bm{0}$. Therefore, when implementing the EM algorithm using equations \eqref{E-step u}-\eqref{M-step nu}, we need to set $\bm{a}=\bm{0}$ everywhere it appears and do not update it in the M-step (equation \eqref{M-step a}). 

Once the parameters $\bm{a}_i$, $\bm{b} $, $\bm{b}_i$, $k_t$ and $k_{i,t}$ have been estimated, the time series modelling and forecasting stages remain unchanged as in \cite{multiLC-2005-Li}. More specifically, the sequence $\{k_t \}$ is modeled by a drifted random walk, and each $\{k_{t,i}\}$ is fitted by a stationary AR(1) process.

\subsection{Common Age-Effect (CAE) Model}
The second multi-population extension we consider is the CAE model proposed by \cite{CAE-2015-Kleinow}. Recall that the ACF model specifies a common factor $b_xk_t$ and a specific factor $b_{x,i}k_{t,i}$ for each population as shown in \eqref{ACF}. By contrast, the CAE model instead assigns each population a common age effect $b_x$ and a specific time trend $k_{t,i}$. Mathematically, for population $i$,
\begin{equation}\label{CAE}
    y_{x,t,i}:=\log(m_{x,t,i})=a_{x,i}+b_xk_{t,i},
\end{equation}
or in the vector form:
\begin{equation}\label{CAE vector}
    \bm{y}_{t,i}:=\log(\bm{m}_{t,i})=\bm{a}_i+\bm{b}k_{t,i}.
\end{equation}
\cite{CAE-2015-Kleinow} demonstrated that the estimation procedure can be treated as a special case of a method called common principal component analysis, which relies on certain specific numerical algorithms for finding the solutions, as noted in \cite{CPCA-1988-Clarkson} for instance. However, it turns out that the estimation can be done with more ease by augmenting the data matrix as follows:
\begin{enumerate}
    \item For each of $i=1,\cdots,I$, take $\hat{\bm{a}}_{i}=\Bar{\bm{y}}_{i}$ and compute the centered log mortality rates at time $t$,  $\Tilde{\bm{y}}_{t,i}=\bm{y}_{t,i}-\Bar{\bm{y}}_{i}$. This is the same as Step 1 for the ACF model.
    \item Construct an augmented data matrix $\Tilde{\bm{Y}}$ by combining the centered log death rates $\Tilde{\bm{y}}_{t,i}$ for each population by column:
    \begin{equation}\label{CAE augmented Y}
        \Tilde{\bm{Y}}=(\underbrace{\Tilde{\bm{y}}_{1,1},\cdots,\Tilde{\bm{y}}_{n,1}}_{\Tilde{\bm{Y}}_1} ,\underbrace{\Tilde{\bm{y}}_{1,2},\cdots,\Tilde{\bm{y}}_{n,2}}_{\Tilde{\bm{Y}}_2} ,\cdots,\underbrace{\Tilde{\bm{y}}_{1,I},\cdots,\Tilde{\bm{y}}_{n,I}}_{\Tilde{\bm{Y}}_I} ),
    \end{equation}
    where $\Tilde{\bm{Y}}_i=(\Tilde{\bm{y}}_{1,i},\cdots,\Tilde{\bm{y}}_{n,i})$ represents the centered log mortality rate matrix for population $i$. 
    \item Apply SVD to the augmented matrix $\Tilde{\bm{Y}}$ to find $\hat{\bm{b}}$.
    \item For each $t=1,\cdots,n$ and $i=1,\cdots,I$, $k_{t,i}$ can be found by \eqref{Lee-Carter k solution}. 
\end{enumerate}

To perform the purposed robust estimation method, we simply replace the SVD in Step 3 with the $t$-PPCA model. Also, similar to the ACF model, the data matrix $\Tilde{\bm{Y}}$ has already been centered, so we need to fix $\bm{a}=\bm{0}$ and do not update it in the M-step using \eqref{M-step a}. After obtaining the estimates of all the parameters, one can choose any time series model that meets the purpose, to fit the time trend sequence $\{k_{t,i}\}$ for each population.


\section{Numerical Analysis}
In this section, we apply the proposed $t$-PPCA method to U.S. mortality data. The data are obtained from Human Mortality Database (HMD), a resource extensively used in actuarial science and demographic research. The death numbers ($D_{x,t}$), exposure to risk ($E_{x,t}$), and central mortality rates ($m_{x,t}$) span all genders and are indexed by single year of age (0-100). For the entirety of the numerical analysis, we employ the standard SVD and Poisson GLM as benchmarks for comparison, with the Poisson GLM method implemented via the ``\textbf{StMoMo}” package in \textbf{R}.

\subsection{Illustration: The United States Mortality Rates and World War II}
We start by applying the proposed $t$-PPCA model to the actual U.S. mortality dataset to demonstrate its robustness in parameter estimation. We investigate two distinct calibration windows: 1940-2019 and 1970-2019. The former window is particularly noteworthy for the inclusion of potential outliers, the most significant of which is World War II. In contrast, the latter window, spanning from 1970 to 2019, does not feature any major historical events that would significantly influence mortality rates. Importantly, our analysis is confined to data up to the year 2019 to deliberately exclude the impact of the Covid-19 pandemic.

The fitting results are presented in Figure 1. For the 1970-2019 calibration window, all three methods yield strikingly similar estimates, as anticipated due to the lack of severe outliers in the post-war period with no major pandemic emerged. Hence, the benefits of robustification are marginal in this context. For the 1940-2019 interval, however, the $t$-PPCA model reveals distinct estimates from the benchmark methods. The $b_x$ estimated via the $t$-PPCA method appears more robust, closely mirroring the counterpart in the 1970-2019 dataset that lacks significant outliers. In contrast, SVD and Poisson GLM methods yield $b_x$ estimates approximating a flat line for ages from 20 to 80, which is a marked deviation. 

\begin{figure}[tp]
    \centering
    \includegraphics[width=16cm]{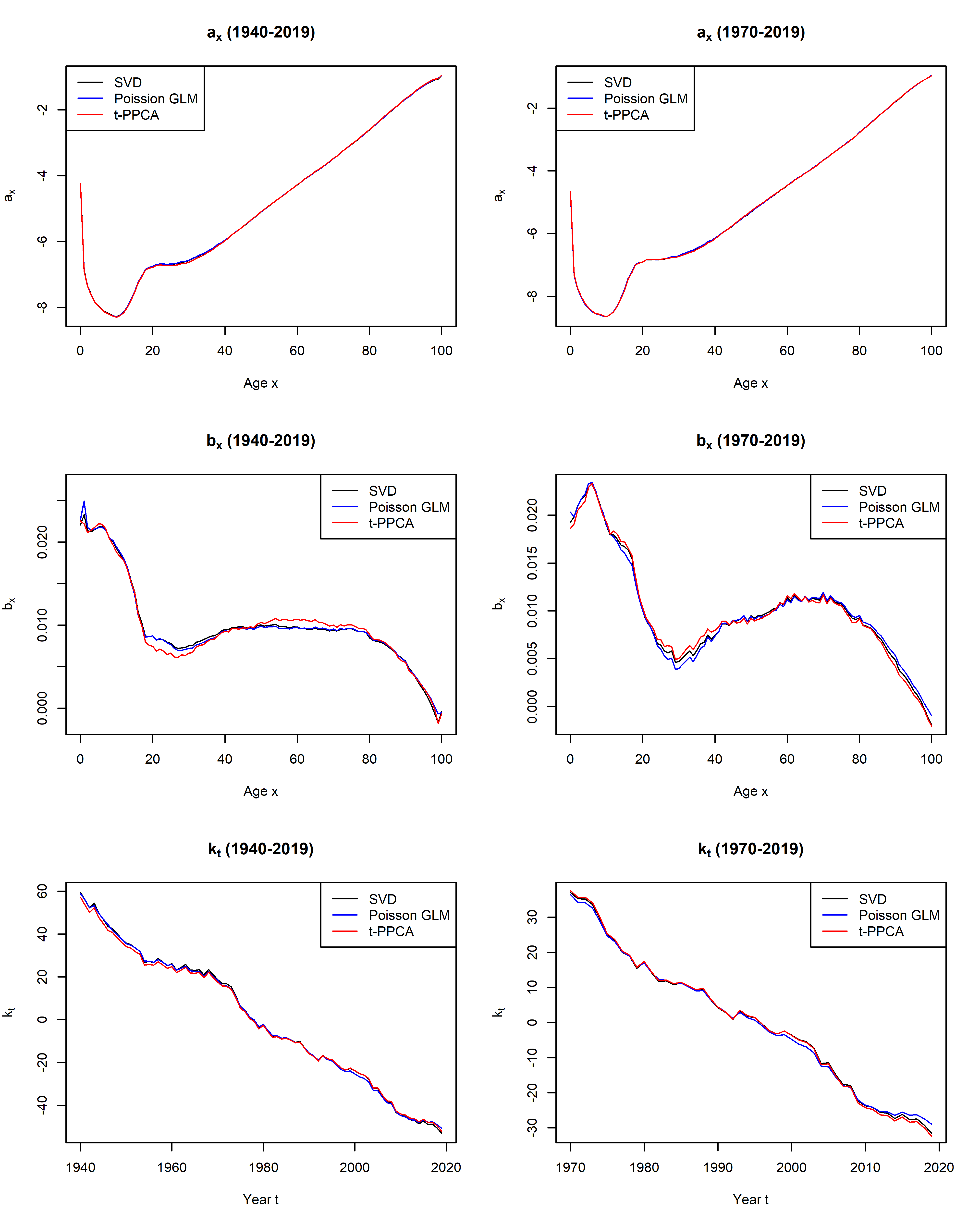}
    \caption{Estimates of $a_x$ (top), $b_x$ (middle) and $k_t$ (bottom) for real U.S. mortality data. Left: 1940-2019; right: 1970-2019.}
\end{figure}

Let us discuss deeper about the difference in the patterns of $b_x$ estimated from the two calibration windows. One key observation is that the benchmark methods tend to overestimate $b_x$ for individuals aged 20-30. This phenomenon becomes evident when we consider the influential outlier events within the time frame of 1940-1969, primarily WWII.

WWII significantly skews the mortality data for young to middle-aged individuals. This phenomenon can be traced back to the characteristics of the American servicemen who participated in the conflict. According to \cite{WWII-1947-Smith}, almost half of the servicemen were under 26 years of age, and 42.6\% were between 26 and 37, with only a scant 7.5\% aged 38 or over. In contrast, the overall male population in 1940 had a more balanced age distribution, with only 29\% under 26 and 32.8\% aged 38 or older. This indicates that the U.S. military predominantly enlisted younger individuals, which resulted in a higher death toll in that specific age range, thereby causing the excess in the early 1940s mortality data. The impact of WWII is evident when considering that it resulted in approximately 405,399 American deaths, but in contrast, the Korean War and the Vietnam War caused approximately 36,574 and 58,220 U.S. deaths, respectively \citep{Wardeath-2020-CRS}. These numbers clearly indicate that WWII was a more dominant outlier, causing a significantly higher loss of life in a relatively short time span compared to the other conflicts. 

Additionally, according to the Centers for Disease Control and Prevention (CDC), the 1940-1969 time frame did not witness any significant pandemics that could affect the mortality rate. It implies that the primary outlier events that skewed mortality were war-related, mainly WWII. To sum up, the inherent characteristics of the young servicemen recruited during WWII have a pronounced impact on the overestimation of $b_x$ for ages 20-30 when applying benchmark methods like SVD and Poisson GLM. 

This case study, while illustrative, does not fully prove the robustness of our proposed method. A more comprehensive and practical simulation study is addressed in Section 5.2, where we quantitatively assess parameter estimation quality across various methods in the presence of outliers via simulation studies.

We conclude Section 5.1 by addressing the influence of initial values on the application of the $t$-PPCA approach to fit a Lee-Carter model. Specifically, the EM algorithm, as detailed in Section 3.3, requires initial values for the parameters $(\bm{a},\bm{b},\sigma^2,\nu)$ of the $t$-PPCA model $\bm{y}_t \sim t_{\nu}(\bm{a},\bm{b}\bm{b}^T+\sigma^2\bm{I})$. It is well-established that the EM algorithm may only find a local maximum of the log-likelihood function, and poorly chosen initial values could lead to sub-optimal MLE of the parameters. Given that the $t$-PPCA model is a modification of the standard PPCA model, which has closed-form MLE and is straightforward to fit, we use the MLE from the standard PPCA model as natural starting values for the shared parameters $(\bm{a},\bm{b},\sigma^2)$, as specified in \eqref{Lee-Carter PPCA solution}. For the additional degrees of freedom parameter $\nu$ in the $t$-PPCA model, we initialize it at $\nu=3$ following \cite{MyPPCA-2022-Guo}. These initial values are maintained throughout Section 5 unless explicitly stated otherwise.

To assess the robustness and stability of the EM algorithm, we conduct sensitivity analyses on the initial values for both calibration windows: 1940-2019 and 1970-2019. For $\bm{a}$ and $\bm{b}$, we consider variations of $\pm 20\%$ from the default initial values, which are derived from the standard PPCA model as described in the last paragraph. For $\sigma^2$, we consider scaling the initial value by factors of 0.1 and 10. For the degrees of freedom $\nu$, apart from the default value, we experiment with both a lower value $\nu=1.5$ and a large value $\nu=20$. In each experiment, only one parameter among $\bm{a}$, $\bm{b}$, $\sigma^2$ and $\nu$ is altered, while the others remain fixed. The sensitivity analysis reveals that the EM algorithm is highly robust to variations in the initial values of $\bm{b}$, $\sigma^2$ and $\nu$ as all tests yield nearly identical MLE with comparable computation times. In contrast, the initial values for $\bm{a}$ do impact the algorithm's computational efficiency, although the MLE obtained are largely consistent. However, this is generally not a concern in practice. This is because $a$ in the Lee-Carter model inherently serves as an indicator of the average level of log mortality rates. Therefore, the recommended initial value $\bar{\bm{y}}$, derived from the standard PPCA as per \eqref{Lee-Carter PPCA solution} gives a natural and reasonable starting point. Given this, there is limited rationale for deliberately choosing alternative initial values for $\bm{a}$.

\subsection{Full Simulation Studies with Hypothetical Pandemic}
In this subsection, we design a series of simulation experiments to comprehensively evaluate the robustness of our proposed $t$-PPCA Lee-Carter model. Without question, Covid-19 is one of the most significant public health events since 2020, causing an unexpectedly high number of deaths in many countries. However, when projecting mortality using a model from the Lee-Carter family for a relatively distant future, say 2050, the years most impacted by Covid-19 should be treated as outliers, and the age-dependent parameters $a_x$ and $b_x$ should be estimated robustly against such outliers. 

The fundamental idea of these simulation studies is to assume a severe pandemic, such as Covid-19, occurred in history and then examine how robust our proposed method performs compared to the SVD and Poisson GLM approaches. The experiments are based on the U.S. mortality data from 1970 to 2019 obtained from HMD. As discussed in Section 5.1, this time period contains few pre-existing extreme war-related and pandemic-related events, this is suitable for simulation to control the artificial outlier effect. 

To introduce hypothetical outliers, we supplement the historical mortality data with the total number of deaths involving Covid-19 in the U.S. in 2020 for all sexes. The Covid-19 data are obtained from the website of the Centers of Disease Control and Prevention (CDC)\footnote{https://data.cdc.gov/NCHS/Provisional-COVID-19-Deaths-by-Sex-and-Age/9bhg-hcku} and are displayed in Table 1. Unlike the other mortality data, which are indexed by single year of age, the Covid-19 mortality data are aggregated into broader age groups.

To distribute these aggregated Covid-19 death counts into individual age groups, we employ an approach based on the ``proportionality hypothesis". This hypothesis suggests that Covid-19 death rates by age are approximately proportional to all-cause mortality rates by age, as supported by existing studies such as \citep{Covid-2020-Cairns,Covid-2023-Cairns}. Essentially, this allows us to use existing empirical death rate distributions to accurately partition the Covid-19 death counts across various age groups. According to the proportionality hypothesis, if a particular age group has a higher all-cause mortality rate, it is reasonable to assume that the Covid-19 mortality rate will be correspondingly higher for that age group. Therefore, we utilize the empirical distribution of total U.S. deaths in 2020 to apportion the aggregated Covid-19 deaths into individual age groups. This method offers a more nuanced and theoretically grounded way of integrating the Covid-19 death data into our analysis.

As an initial illustration, we consider a scenario where a ``hypothetical Covid-19" pandemic occurred for three years from 1970-1972. In terms of data, we incorporate the redistributed Covid-19 deaths (U.S. 2020) into the total deaths in 1970, 1971, and 1972. Although the number of deaths from a pandemic will vary year to year in reality, we assume it remains constant in this simulation for simplicity's sake, as our goal is to generate outliers rather than focusing on their precise values. For each of the SVD (Figure 2, top), Poisson GLM (Figure 2, middle), and the proposed robust $t$-PPCA (Figure 2, bottom), we estimate parameters with and without the added outliers and compare the results.
\begin{table}[tp]
\centering
\begin{tabular}{ c| c c c c c c }
\hline \hline
Age Group & $<1$ & 1-4 & 5-14 & 15-24 & 25-34 &35-44  \\

Deaths & 52 & 25 & 68 & 615 & 2,621 & 6,785   \\
\hline
Age Group & 45-54 & 55-64 & 65-74 & 75-84 & $>85$ & N/A  \\

Deaths & 18,327 & 45,572 & 82,286 & 106,259 & 122,820 & N/A   \\
\hline \hline
\end{tabular}
\caption{U.S. Covid-19 Deaths in 2020}
\end{table}

In this experiment, the estimates of $a_x$ are essentially unaffected when adding outliers for all three methods as seen from Figure 2, which is consistent with Figure 1 and implies that $\Hat{a}_x$ seems to inherently be robust to outliers. This observation is later formalized in Table 2. On the other hand, both the SVD and Poisson GLM methods produce highly fragile estimates for $b_x$ in the presence of outliers, which is visible in Figure 2. As expected, $b_x$ is overestimated for older ages, as most of the deaths in the added outlier (hypothetical Covid-19) are among the older population (as indicated in Table 1) and occur at the beginning of the experimental period (1970-1972). Given that $b_x$ represents the sensitivity of the death rates with respect to the time index $k_t$, a positive outlier at the beginning will inflate the overall sensitivity $b_x$. Conversely, our proposed $t$-PPCA approach provides extremely robust estimates across all ages $x=0,\cdots,100$ for $b_x$, even when outliers are inserted.

It is worth studying the plots for $\Hat{b}_x$ for SVD and Poisson GLM in more detail. As seen from Figure 2, it may appear that the SVD and Poisson GLM underestimate $b_x$ for young age groups in the simulated scenario, but this interpretation is not accurate. Note that the death number in the outlier generally increases with age (and is always positive), so $b_x$ should theoretically be overestimated for all ages and to a growing extent as age increases. However, the imposed parameter constraint $\sum_x \hat{b}_x=1$ re-scales the overall overestimation, making the less overestimated $b_x$ part (younger age groups) appear as an ``underestimation".

As for the time index $k_t$, as observed from Figure 2, all methods yield significantly distorted estimates at $t=1970,1971,1972$ (the years containing outliers), which is intuitive. For SVD and $t$-PPCA, $\Hat{k}_t$ is obtained from the death number matching $D_t=\sum_x N_{x,t}\cdot e^{\hat{a}_{x}+\hat{b}_{x}k_t}$ after obtaining $\hat{a}_x$ and $\hat{b}_x$. As a result, a positive outlier with a larger $D_t$ will naturally produce a positively deviated $\hat{k}_t$, regardless of the quality of $\hat{a}_x$ and $\Hat{b}_x$ estimation. Similarly, an abnormal $D_t$ of the outliers will be directly used in the Poisson likelihood structure, which consequently leads to poor $\hat{k}_t$ estimation. Despite the distorted estimates $k_t$ for the outliers, our proposed $t$-PPCA outperforms the traditional SVD and Poisson GLM. The $t$-PPCA provides highly robust estimates for $k_t$ where $t$ is not an outlier point (as depicted in Figure 2 bottom for the years 1973-2019), indicating that the outlier effect does not propagate significantly beyond the initially affected time region. However, the estimated $k_t$ in the ``normal time region" obtained from SVD and Poisson GLM (as seen in Figure 2 top/middle for the years 1973-2000) is still influenced; this indicates that the outlier effect propagates much further.

\begin{figure}[tp]
    \centering
    \includegraphics[width=16cm]{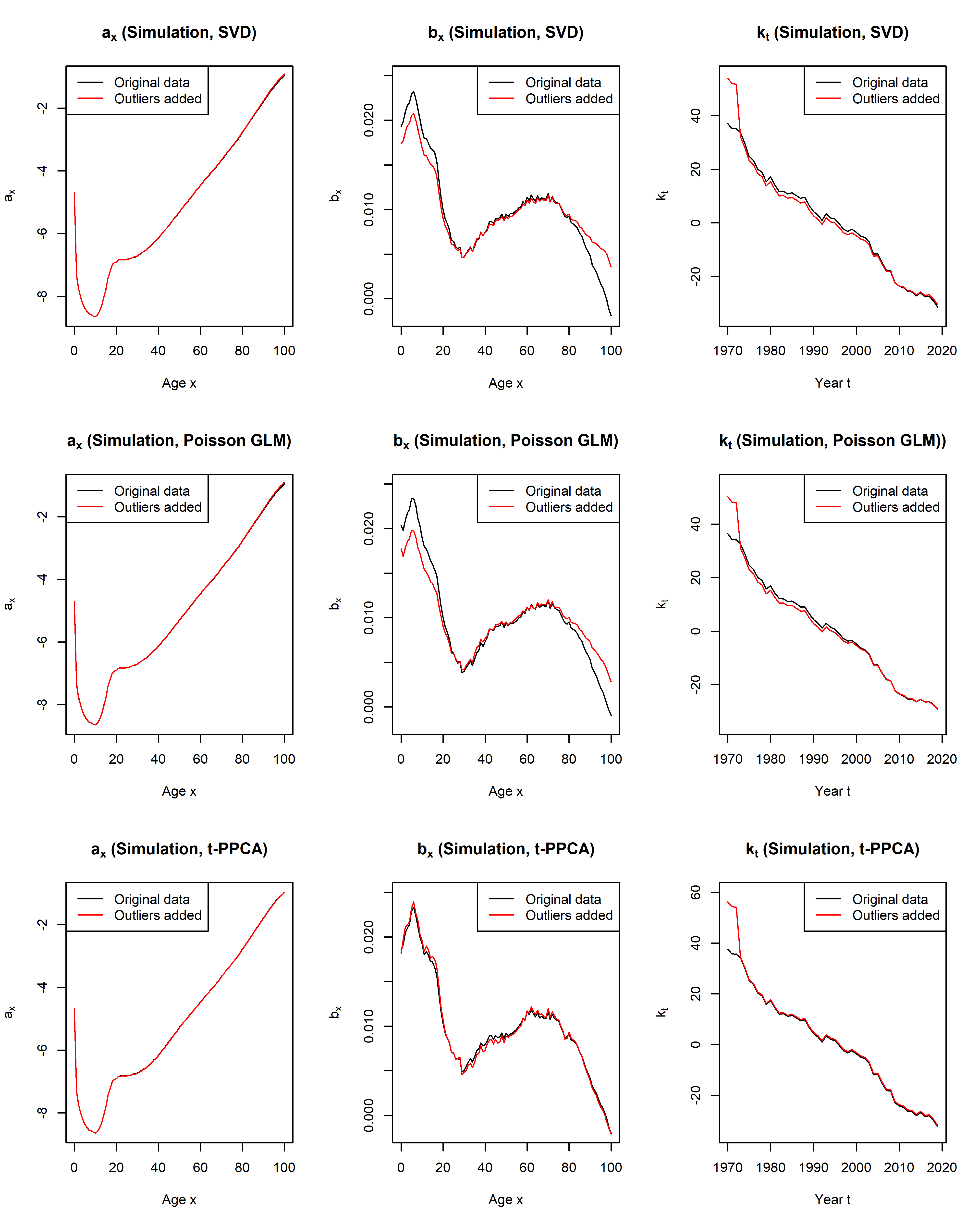}
    \caption{Estimated parameters for the simulated U.S. mortality data from year 1970 to 2019, with Covid-19 outliers added to year 1970-1972: SVD (top); Poisson GLM (middle); $t$-PPCA (bottom).}
\end{figure}

To further evaluate the performance of the three estimating methods, we consider three different scenarios where the ``hypothetical Covid-19" (added outlier) lasted for one, three, or five years in history. We conduct a set of experiments for each scenario, with the outlier data following Table 1, which outlines the Covid-19 deaths in the U.S. in 2020. For example, in the set of experiments where the outlier lasts for one year, we insert this one-year outlier into each of the years 1970-2019, resulting in a total of $N=50$ experiments. For another set of experiments where the outlier lasts for three years, we insert the outlier into the years 1970-1972, 1971-1973, 1972-1974,$\cdots$, 2017-2019, which results in a total of $N=48$ experiments. The scenario for five-year outliers is similar to the three-year case and involves $N=46$ experiments. The aim of these simulations is to examine the robustness of the three methods, assuming a ``hypothetical Covid-19" event in history, lasting one, three, or five years, respectively.
\begin{table}[tp]
\centering
\begin{tabular}{c c|c c}
\hline \hline
Outliers   & Method  &  Average RMAE $(\hat{a})$  & Average RRMSE $(\hat{a})$ \\ \hline
\multirow{3}{*}{1-year}   
& SVD  & $0.0010 $  &$0.0018$ \\ 
& Poisson GLM & $0.0013$  & $0.0027$ \\ 
& $t$-PPCA  & $0.0006$  &$0.0008$  \\ \hline
\multirow{3}{*}{3-year}
& SVD  & $0.0029$ & $0.0054$   \\ 
& Poisson GLM &$0.0038$  &$0.0079$   \\ 
& $t$-PPCA& $0.0017$  &$0.0023$   \\ \hline
\multirow{3}{*}{5-year}
& SVD  & $0.0048$  & $0.0089$   \\ 
& Poisson GLM & $0.0060$  &$0.0124$  \\ 
& $t$-PPCA & $0.0028$  & $0.0038$ \\ \hline \hline
\\
\hline \hline

Outliers   & Method  &  Average RMAE $(\Hat{b})$  & Average RRMSE $(\Hat{b})$  \\ \hline
\multirow{3}{*}{1-year}   
& SVD  & $0.0448 $ &$0.1890$ \\  
& Poisson GLM & $0.0705$ & $0.3001$\\
& $t$-PPCA  & $0.0170$ &$0.0472$ \\ \hline
\multirow{3}{*}{3-year}
& SVD  & $0.1220$ & $0.5120$  \\ 
& Poisson GLM &$0.1919$ &$0.8200$ \\ 
& $t$-PPCA& $0.0479$ &$0.1326$ \\ \hline
\multirow{3}{*}{5-year}
& SVD  & $0.1851$ & $0.7730$ \\ 
& Poisson GLM & $0.2909$ &$1.2463$ \\ 
& $t$-PPCA & $0.0746$ & $0.2028$\\ \hline \hline\\
\hline \hline

Outliers   & Method  &  Average RMAE $(\hat{k})$  & Average RRMSE $(\hat{k})$ \\ \hline
\multirow{3}{*}{1-year}   
& SVD  & $0.0725 $ &$0.1647$ \\ 
& Poisson GLM & $0.0532$  & $0.1027$\\
& $t$-PPCA  & $0.0379$ &$0.0905$ \\ \hline
\multirow{3}{*}{3-year}
& SVD  & $0.2155$ & $0.4878$  \\
& Poisson GLM &$0.1640$ &$0.3149$ \\
& $t$-PPCA& $0.1240$ &$0.2934$ \\ \hline
\multirow{3}{*}{5-year}
& SVD  & $0.3554$ & $0.8002$ \\ 
& Poisson GLM & $0.2711$ &$0.5178$ \\
& $t$-PPCA & $0.2187$  & $0.5135$\\ \hline \hline
\end{tabular}
\caption{Estimation errors of $\hat{a}_x$, $\hat{b}_x$ and $\hat{k}_t$ under different settings of hypothetical pandemic.}
\end{table}

In each experimental setup across three different scenarios for outliers, we obtain estimates for the parameters $a_x$, $b_x$, and $k_t$, denoted as $\hat{a}_x$, $\hat{b}_x$ and $\hat{k}_t$, respectively. To evaluate the performance of various estimation methods, we employ two key performance metrics: Relative Mean Absolute Error (RMAE) and Relative Root Mean Square Error (RRMSE). Notably, RRMSE imposes a stricter penalty on larger errors compared to RMAE. These metrics are chosen because they provide a relative measure of estimation errors, a critical feature given the diverse magnitudes of different parameters. The metrics are formally defined as follows:
\begin{equation}\label{Error metrics a}
    \text{RMAE}(\hat{\bm{a}})=\frac{1}{p}\sum_{x=x_1}^{x_p}\left|\frac{\hat{a}_x-\hat{a}_x^{(0)}}{\hat{a}_x^{(0)}} \right|,\quad \text{RRMSE}(\hat{\bm{a}})=\sqrt{\frac{1}{p}\sum_{x=x_1}^{x_p}\left(\frac{\hat{a}_x-\hat{a}_x^{(0)}}{\hat{a}_x^{(0)}} \right)^2} , 
\end{equation}
\begin{equation}\label{Error metrics b}
    \text{RMAE}(\hat{\bm{b}})=\frac{1}{p}\sum_{x=x_1}^{x_p}\left|\frac{\hat{b}_x-\hat{b}_x^{(0)}}{\hat{b}_x^{(0)}} \right|,\quad \text{RRMSE}(\hat{\bm{b}})=\sqrt{\frac{1}{p}\sum_{x=x_1}^{x_p}\left(\frac{\hat{b}_x-\hat{b}_x^{(0)}}{\hat{b}_x^{(0)}} \right)^2}, 
\end{equation}
\begin{equation}\label{Error metrics k}
    \qquad \text{RMAE}(\hat{\bm{k}})=\frac{1}{|\mathcal{T}|}\sum_{t\in \mathcal{T}}\left|\frac{\hat{k}_t-\hat{k}_t^{(0)}}{\hat{k}_t^{(0)}} \right|,\quad
    \text{RRMSE}(\hat{\bm{k}})=\sqrt{\frac{1}{|\mathcal{T}|}\sum_{t\in \mathcal{T}}\left(\frac{\hat{k}_t-\hat{k}_t^{(0)}}{\hat{k}_t^{(0)}} \right)^2} ,
\end{equation}
where $\hat{a}_x^{(0)}, \hat{b}_x^{(0)}, \hat{k}_t^{(0)}$ are the baseline estimates obtained in the absence of outliers, serving as a fixed reference in all experiments within each scenario.

In \eqref{Error metrics k}, the set $\mathcal{T}$ represents the years under investigation, specifically excluding those that contain hypothetical outliers. The cardinality of $\mathcal{T}$ is denoted as $|\mathcal{T}|$. More precisely, $\mathcal{T}$ is defined as:
\begin{equation}\label{k no outlier subset}
    \mathcal{T} = \{ t \in \{1970, 1971, \cdots, 2019\}: t \notin \mathcal{O} \},
\end{equation}
where $\mathcal{O}$ represents years with artificially introduced outliers. For example, in a setup with outliers occurring in the years 1973, 1974, and 1975, we would have $ \mathcal{O} = \{1973, 1974, 1975\}$ and $ \mathcal{T} = \{1970, 1971, 1972, 1976, 1977, \cdots, 2019\} $, with $|\mathcal{T}| = 50 - 3 = 47 $. We define $\mathcal{T}$ in this manner to exclude years affected by outliers, as the parameter $\hat{k}_t$ is inherently distorted for these years, a phenomenon already illustrated in Figure 2. The primary objective is to assess the robustness of $\hat{k}_t$ over time periods in time frames without artificially introduced outliers.

In each experiment, the RMAE and RRMSE values for $\hat{a}_x$ and $\hat{b}_x$ are calculated by taking the average over all ages, ranging from $x_1=0$ to $x_p=100$. Similarly, the RMAE and RRMSE for $\hat{k}_t$ are averaged across the set $\mathcal{T}$, as previously defined. Subsequently, the mean RMAE and RRMSE are calculated across all $N$ experiments within each outlier scenario. Here, $N$ takes the values 50, 48, or 46, corresponding to the different outlier settings described earlier. The summarized results are presented in Table 2.

Firstly, from the top part of Table 2, the estimates of $a_x$ are highly robust for all scenarios, with an MAPE consistently smaller than $1\%$, even when a severe 5-year outlier is introduced. This phenomenon is observed across all three methods, so our $t$-PPCA method is not particularly attractive for estimating $a_x$. Instead, our primary focus is on the performance of estimating the sensitivity index $b_x$. Under all three outlier scenarios, $\hat{b}_x$ obtained from the Poisson GLM is the most vulnerable, followed by the SVD, while $t$-PPCA yields the most robust estimates, improving the average MAPE by at least $60\%$ compared to the SVD and Poisson GLM. It is also worth noting that $t$-PPCA results in even greater improvement of RMSPE, indicating that our proposed method effectively controls estimation errors of a large magnitude.

With respect to the estimation of the time index $k_t$, $t$-PPCA also reduces the MAPE and RMSPE, yielding a more robust $\Hat{k}_t$ and mitigating the outlier effects in the second-stage time series modeling. Note that although the improvement is relatively conservative compared to that of $b_x$, this should not be a concern when adopting our method to estimate parameters. One benefit of $t$-PPCA is that it solely focuses on the estimation stage, and can be naturally combined with any existing time series model in the forecasting stage without any conflict.


\section{Further Comments and Conclusion}
This paper proposes a systematic approach for estimating the parameters of the Lee-Carter model. Outliers arise naturally in mortality data and their influences are expected to have reduced influence when making long-term mortality projections. Leveraging the equivalence between the traditional SVD approach and the PPCA formulation, we propose a modified PPCA model based on the multivariate $t$-distribution to robustify the estimation quality in the presence of outliers. We have designed a computationally efficient EM algorithm for implementation, and residual bootstrap can be utilized to examine parameter uncertainties. Through empirical studies and various simulation experiments, we demonstrate that the proposed $t$-PPCA model significantly enhances parameter robustness, particularly for $b_x$.

The current literature predominantly focuses on the outlier analysis of the Lee-Carter model's time series modeling step for the time index $k_t$, assuming the sensitivity parameter $b_x$ derived from the estimation stage is accurate. This assumption could lead to skewed mortality projections, even if a convincing model for $k_t$ is applied. Our proposed $t$-PPCA model is primarily aimed at the estimation stage, with the key objective of providing a robust estimate for $b_x$. Any suitable time series model for $k_t$ can be seamlessly integrated with the $t$-PPCA estimation procedure to produce robust mortality projections that account for outliers.

Lastly, we would like to comment on the appropriateness of applying robust methods in mortality modeling. As discussed in the literature \cite{Stoch-2002-Chan, OutlierLC-2005-Li, OutlierLC-2007-Li}, if the goal is to capture the long-term mortality trend, then robust models are favored, as they can diminish the effect of outliers. Conversely, when the quantity of interest is intimately linked to extreme stochastic fluctuations, such as the highest attained age, the existence and impact of outliers should be preserved.

\section*{Appendix A: Some Properties of the Multivariate Normal Distributions and Multivariate $t$-Distributions}
\setcounter{equation}{0}
\renewcommand\theequation{A.\arabic{equation}}

In this appendix, we present some useful properties of the multivariate normal distributions and multivariate $t$-distributions, which are referred multiple times in the main chapters and Appendix B.
We present the results in somewhat simplified forms which sufficient for our purpose. All the results and their generalized forms can be found in many classical machine learning textbooks, for example, \cite{PRML-2006-Bishop}.

\subsection*{Results of Normal-Normal Hierarchy}
The first set of results consider the multivariate normal distributions where the mean parameter is also normally distributed. Let $\bm{y}$ be a $p$-dimensional random vector with the following hierarchical structure:
\begin{equation}\label{Normal-Normal hierarchy}
    \bm{y}|z\sim \mathcal{N}(\bm{a}+\bm{b}z,\bm{\Sigma}), \quad z\sim \mathcal{N}(0,\tau),
\end{equation}
where $(\bm{a},\bm{b},\bm{\Sigma},\tau)$ are fixed parameters, and $\bm{a}+\bm{b}z$ is a linear transformation of the latent random variable $z$. Then, we have the following results:
\begin{equation}\label{Normal-Normal marginal}
    \bm{y}\sim \mathcal{N}(\bm{a},\bm{b}\bm{b}^T\tau+\bm{\Sigma}),
\end{equation}
\begin{equation}\label{Normal-Normal conditional}
    z|\bm{y}\sim \mathcal{N}((\tau^{-1}+\bm{b}^T\bm{\Sigma}^{-1}\bm{b})^{-1}\bm{b}^T\bm{\Sigma}^{-1}(\bm{y}-\bm{a}),(\tau^{-1}+\bm{b}^T\bm{\Sigma}^{-1}\bm{b})^{-1}).
\end{equation}
The results above show that the resulting marginal distribution of $\bm{y}$ and the posterior distribution $z|\bm{y}$ preserve normal.

\subsection*{Results of Normal-Gamma Hierarchy}
The second set of results consider the multivariate normal distributions where the covariance matrix is also inversely proportional to a Gamma random variable. Let $\bm{y}$ be a $p$-dimensional random vector with the following hierarchical structure:
\begin{equation}\label{Normal-Gamma hierarchy}
    \bm{y}|u \sim \mathcal{N} \left(\bm{\mu},\frac{\bm{\Sigma}}{u}\right),\ u \sim \mathrm{Gamma}\left(\frac{\nu}{2},\frac{\nu}{2}\right),
\end{equation}
where $(\bm{\mu},\bm{\Sigma},\nu)$ are fixed parameters. It is worth noting that $\mathrm{Gamma}\left(\nu/2,\nu/2 \right)$ is equivalent to the chi-square distribution $\chi_{\nu}^2$ with $\nu$ as the degrees of freedom. Then, we have the following important result:
\begin{equation}\label{Normal-Gamma marginal}
    \bm{y}\sim t_{\nu}(\bm{\mu},\bm{\Sigma}),
\end{equation}
with the density function defined in \eqref{Multivariate t pdf}. It shows that a multivariate $t$-distribution can be interpreted as a infinite mixture of normal multivariate normal distributions. This property builds up the foundation to find the MLE of the multivariate $t$-distributions, via the EM algorithm.

Another result is a special case of the so-called normal-gamma conjugacy, which is commonly used in Bayesian inference. It shows that the posterior distribution of $u$ preserves Gamma under the normal likelihood:
\begin{equation}\label{Normal-Gamma conditional}
    u|\bm{y} \sim \text{Gamma}\left(\frac{\nu+p}{2}, \frac{\nu+(\bm{y}-\boldsymbol{\mu})^{T} \mathbf{\Sigma}^{-1}(\bm{y}-\boldsymbol{\mu})}{2}\right).
\end{equation}

\section*{Appendix B: Derivation of the EM Algorithm}
\setcounter{equation}{0}
\renewcommand\theequation{B.\arabic{equation}}
This appendix presents the details of deriving all the updating formulas from \eqref{EM complete likelihood expectation} to \eqref{M-step nu} in the EM algorithm as discussed in Section 3.3. The derivations are based on the hierarchical structure of the proposed $t$-PPCA model, as shown in \eqref{Lee-Carter t-PPCA mixture Gaussian representation}.

\subsection*{The Complete Log-Likelihood}
We first present more details of deriving the complete log-likelihood \eqref{EM complete likelihood} and its conditional expectation \eqref{EM complete likelihood expectation} in the E-step. The derivation relies on the probability density functions of the distributions involved in the $t$-PPCA hierarchical structure \eqref{Lee-Carter t-PPCA mixture Gaussian representation}:
\begin{equation}\label{pdf y|z,u}
    p(\bm{y}_t|z_t,u_t)=\left((2\pi)^p \cdot  \left| \frac{\sigma^2 \bm{I}}{u_t}\right| \right)^{-1/2} \cdot \exp \left(-\frac{1}{2}\left(\bm{y}_t-\bm{a}-\bm{b}z_t \right)^T\cdot \left(\frac{\sigma^2 \bm{I}}{u_t} \right)^{-1} \cdot \left(\bm{y}_t-\bm{a}-\bm{b}z_t \right) \right),
\end{equation}

\begin{equation}\label{pdf z|u}
    p(z_t|u_t)=\left(\frac{2\pi}{u_t}\right)^{-1/2} \cdot \exp \left(-\frac{z_t^2}{2u_t^{-1}} \right),
\end{equation}

\begin{equation}\label{pdf u}
    p(u_t)=\frac{(\nu/2)^{\nu/2}}{\Gamma(\nu/2)} u_t^{(\nu/2)-1}\cdot \exp\left(-\frac{\nu}{2}u_t \right).
\end{equation}

Then, the complete log-likelihood can be computed as follows: 
\begin{align}
    L_c &= \sum_{t=1}^n \log [p(\bm{y}_t, z_t, u_t)]=\sum_{t=1}^n \log[p(\bm{y}_t|z_t,u_t)p(z_t|u_t)p(u_t)]\nonumber \\
    &= -\sum_{t=1}^n \bigg [\frac{1}{2}\log \left|\frac{\sigma^2\bm{I}}{u_t} \right| +\frac{1}{2}(\bm{y}_t-\bm{a}-\bm{b}z_t)^T\cdot \frac{u_t\bm{I}}{\sigma^2}\cdot (\bm{y}_t-\bm{a}-\bm{b}z_t)\nonumber \\
    &\quad +\frac{1}{2}\log \left(\frac{1}{u_t}\right)+\frac{1}{2}u_tz_t^2-\frac{\nu}{2} \left(\log \frac{\nu}{2}+\log u_{t}- u_{t} \right)+\log \Gamma \left(\frac{\nu}{2} \right)\bigg]+\mathrm{const}. \nonumber \\
    &= -\sum_{t=1}^n \bigg [\frac{p}{2}\log \sigma^2 +\frac{u_t}{2\sigma^2}(\bm{y}_t-\bm{a})^T(\bm{y}_t-\bm{a})-\frac{1}{\sigma^2}(u_tz_t)\bm{b}^T(\bm{y}_t-\bm{a})\nonumber \\
    &\quad + \frac{1}{2\sigma^2}\bm{b}^T\bm{b}u_tz_t^2 -\frac{\nu}{2} \left(\log \frac{\nu}{2}+\log u_{t}- u_{t} \right)+\log \Gamma \left(\frac{\nu}{2} \right)\bigg]+\mathrm{const}..
\end{align}
Note that the parameters to be estimated are $(\bm{a},\bm{b},\sigma^2,\nu)$, so any terms that do not involve the parameters can be treated as constants. Then, taking the conditional expectations on both sides with respect to the observed data $\bm{y}_t$'s, we obtain \eqref{EM complete likelihood expectation}, that is, $\langle L_c \rangle$.

\subsection*{The Expectations in the E-Step}
Next, we present the detailed derivations for the posterior expectations \eqref{E-step u}-\eqref{E-step uz2} in the E-step in Section 3.3. 

\subsubsection*{1. $ \langle u_{t}\rangle$ \eqref{E-step u}}

First, from \eqref{Lee-Carter t-PPCA mixture Gaussian representation}, we have $\bm{y}_t|u_t \sim \mathcal{N} \left(\bm{a},(\bm{b}\bm{b}^T+\sigma^2\bm{I})/u_t\right)$, as shown in \eqref{Lee-Carter t-PPCA p(y|u)}, and $u_t \sim \text{Ga}\left( \nu/2,\nu/2 \right)$. By the conjugacy between normal likelihood and gamma prior, we obtain
\begin{equation}\label{u|y}
    u_t|\bm{y}_t\sim \text{Ga}\left(\frac{p+\nu}{2},\frac{(\bm{y}_t-\bm{a})^T(\bm{b}\bm{b}^T+\sigma^2\bm{I})^{-1}(\bm{y}_t-\bm{a})+\nu}{2} \right),
\end{equation}
which immediately implies \eqref{E-step u}:
\begin{equation}
    \langle u_{t}\rangle=\E[u_t|\bm{y}_t]=\frac{\nu+p}{\nu+(\bm{y}_t-\bm{a})^T(\bm{b}\bm{b}^T+\sigma^2\bm{I})^{-1}(\bm{y}_t-\bm{a})}
\end{equation}

\subsubsection*{2. $ \langle \log u_{t}\rangle$ \eqref{E-step log u}}
Following \eqref{u|y}, we can then derive $\langle \log u_t\rangle$, that is, $\E[\log u_t|\bm{y}_t]$. Consider a random variable $Y=\log X$, where $X\sim \text{Ga}(\alpha,\beta)$. Its first moment can be easily derived through the moment generating function:
\begin{align}
    M_Y(t)&=\E[e^{Yt}]=\E[X^t]=\frac{\Gamma(\alpha+t)}{\Gamma(\alpha)}\cdot \beta^{-t}\nonumber \\
\Longrightarrow \E[Y]&=M^{\prime}_Y(t)|_{t=0}=\frac{1}{\Gamma(\alpha)}\left[\Gamma^{\prime}(\alpha+t)\beta^{-t}-\Gamma(\alpha+t)\beta^{-t}\log \beta \right]\big|_{t=0}\nonumber \\
    &=\psi(\alpha) -\log \beta,
\end{align}
where $\psi(\cdot)=\Gamma(\cdot)/\Gamma^{\prime}(\cdot)$ is the digamma function. Substituting $\alpha$ and $\beta$ by the corresponding parameters in \eqref{u|y} immediately gives:
\begin{equation}
    \langle \log u_{t}\rangle=\E[\log u_t|\bm{y}_t]= \psi \left(\frac{\nu+p}{2} \right)-\log \left(\frac{\nu+(\bm{y}_t-\bm{a})^T(\bm{b}\bm{b}^T+\sigma^2\bm{I})^{-1}(\bm{y}_t-\bm{a})}{2} \right).
\end{equation}

\subsubsection*{3. $ \langle z_{t}\rangle$ \eqref{E-step z}}
As presented in \eqref{Lee-Carter t-PPCA mixture Gaussian representation}, we have $\bm{y}_t|z_t,u_t \sim  \mathcal{N} \left(\bm{a}+\bm{b}z_t,\sigma^2\bm{I}/u_t\right)$ and $z_t|u_t \sim \mathcal{N}(0,1/u_t)$. To derive \eqref{E-step z}, we apply result \eqref{Normal-Normal conditional} for the conditioning of multivariate normal distributions, and we can obtain
\begin{equation}\label{z|y,u}
    p(z_t|\bm{y}_t,u_t)\propto p(\bm{y}_t|z_t,u_t)p(z_t|u_t)\sim \mathcal{N}\left((\bm{b}^T\bm{b}+\sigma^2)^{-1}\bm{b}^T(\bm{y}_t-\bm{a}),\frac{\sigma^2(\bm{b}^T\bm{b}+\sigma^2)^{-1}}{u_t}  \right).
\end{equation}
Next, since $u_t\sim \text{Ga}(\nu/2,\nu/2)$, applying the scale mixture Gaussian representation of multivariate $t$-distributions \eqref{Normal-Gamma hierarchy} and \eqref{Normal-Gamma marginal} immediately applies
\begin{equation}\label{z|y}
    z_t|\bm{y}_t\sim t_{\nu}\left((\bm{b}^T\bm{b}+\sigma^2)^{-1}\bm{b}^T(\bm{y}_t-\bm{a}),\sigma^2(\bm{b}^T\bm{b}+\sigma^2)^{-1}  \right).
\end{equation}
It immediately gives \eqref{E-step z}:
\begin{equation}
    \langle z_t \rangle=\E[z_t|\bm{y_t}]=(\bm{b}^T\bm{b}+\sigma^2)^{-1}\bm{b}^T(\bm{y}_t-\bm{a}).
\end{equation}

\subsubsection*{4. $ \langle u_tz_{t}\rangle$ \eqref{E-step uz}}

By the law of total expectations:
\begin{equation}
    \langle u_tz_t \rangle=\E[u_tz_t|\bm{y}_t] =\E[\E[u_tz_t|u_t,\bm{y}_t]|\bm{y}_t]=\E[u_t|\bm{y}_t]\cdot \E[z_t|u_t,\bm{y}_t]=\langle u_t\rangle \langle z_{t}\rangle.
\end{equation}
The second last step is implied by the observation from \eqref{z|y,u} that $\E[z_t|u_t,\bm{y}_t]$ depends on $\bm{y}_t$ only, but not $z_t$ and $u_t$. The last step uses $\E[z_t|u_t,\bm{y}_t]=\E[z_t|\bm{y}_t]$, which can be noticed from \eqref{z|y,u} and \eqref{z|y}.

\subsubsection*{5. $ \langle u_tz^2_{t}\rangle$ \eqref{E-step uz2}}
Similarly, by the law of total expectations:
\begin{align}
    \langle u_tz_t^2 \rangle&= \E[\E[u_tz_t^2|u_t,\bm{y}_t]|\bm{y}_t]\nonumber \\
    &=\E[u_t((\E[z_t|u_t,\bm{y}_t])^2+\V(z_t|u_t,\bm{y}_t))|\bm{y}_t]\nonumber \\
    &=\E[u_t\cdot \V(z_t|u_t,\bm{y}_t)|\bm{y}_t]+\E[u_t|\bm{y}_t]\cdot (\E[z_t|u_t,\bm{y}_t])^2  \nonumber \\
    &=\E[u_t\cdot \V(z_t|u_t,\bm{y}_t)|\bm{y}_t]+\langle u_{t}\rangle \langle z_{t}\rangle^2 \nonumber\\ 
    &=\E \left[u_t\cdot \frac{\sigma^2(\bm{b}^T\bm{b}+\sigma^2)^{-1}}{u_t}\Big|\bm{y}_t\right]+\langle u_{t}\rangle \langle z_{t}\rangle^2 \nonumber\\ 
    &=\sigma^2(\bm{b}^T\bm{b}+\sigma^2)^{-1}+\langle u_{t}\rangle \langle z_{t}\rangle^2.
\end{align}
Similar to the arguments above, the third line again uses the observation that $\E[z_t|u_t,\bm{y}_t]$ depends on $\bm{y}_t$ only, and the fourth line again uses $\E[z_t|u_t,\bm{y}_t]=\E[z_t|\bm{y}_t]$. The fifth line is by $\V(z_t|u_t,\bm{y}_t)=\sigma^2(\bm{b}^T\bm{b}+\sigma^2)^{-1}/u_t$ from \eqref{z|y,u}.

\subsection*{The Updating Formulas in the M-Step}
In this part, we provide more details of deriving the updating formulas \eqref{M-step a}-\eqref{M-step nu} in the M-step, which are obtained by setting all the first order derivatives of $\langle L_c \rangle$ \eqref{EM complete likelihood expectation} with respect to each of the $\bm{a}$, $\bm{b}$, $\sigma^2$ and $\nu$ to 0. 

\begin{gather}
    \frac{\partial \langle L_c \rangle}{\partial \bm{a}}=-\sum_{t=1}^n \left[(-2)\cdot \frac{\langle u_t\rangle}{2\sigma^2}(\bm{y}_t-\bm{a})+\frac{1}{\sigma^2}\langle u_tz_t \rangle \bm{b} \right]=0\\
    \Longrightarrow \quad  \bm{a}= \frac{\sum_{t=1}^{n}\langle u_{t}\rangle \bm{y}_t-\bm{b}\langle u_tz_t\rangle  }{\sum_{t=1}^{n}\langle u_{t} \rangle}=\frac{\sum_{t=1}^{n}\langle u_{t}\rangle \left( \bm{y}_t-\bm{b}\langle z_t\rangle \right) }{\sum_{t=1}^{n}\langle u_{t} \rangle},
\end{gather}

\begin{gather}
    \frac{\partial \langle L_c \rangle}{\partial \bm{b}}=-\sum_{t=1}^n \left[-\frac{1}{\sigma^2}\langle u_tz_t \rangle (\bm{y}_t-a)+2\cdot \frac{1}{2\sigma^2}\langle u_tz_t^2\rangle \bm{b} \right]=0\\
    \Longrightarrow \quad  \bm{b}= \left[\sum_{t=1}^{n}\langle u_{t}z_t^2 \rangle \right]^{-1}\left[\sum_{t=1}^{n}(\bm{y}_t-\bm{a})\langle u_{t}z_t \rangle \right],
\end{gather}

\begin{gather}
    \frac{\partial \langle L_c \rangle}{\partial \sigma^2}=-\sum_{t=1}^n \left[\frac{p}{2\sigma^2}-\frac{\langle u_t \rangle}{2\sigma^4}(\bm{y}_t-\bm{a})^T(\bm{y}_t-\bm{a})+\frac{1}{\sigma^4}\langle u_tz_t \rangle \bm{b}^T(\bm{y}_t-\bm{a})-\frac{1}{2\sigma^4}\bm{b}^T\bm{b}\langle u_tz_t^2\rangle \right]=0\\
    \Longrightarrow \quad  \sigma^2= \frac{1}{np}\sum_{t=1}^{n}\Big[\langle u_{t}\rangle (\bm{y}_t-\bm{a})^T (\bm{y}_t-\bm{a})-2\langle u_{t}z_t \rangle \bm{b}^T (\bm{y}_t-\bm{a})+\bm{b}^T\bm{b}\langle u_{t}z_t^2\rangle \Big],
\end{gather}

\begin{gather}
    \frac{\partial \langle L_c \rangle}{\partial \nu}=-\sum_{t=1}^n \left[-\frac{1}{2}\left(\log \frac{\nu}{2}+\langle \log u_t\rangle -\langle u_t \rangle \right)-\frac{1}{2}+\frac{1}{2}\frac{\Gamma^{\prime}\left(\nu/2\right)}{\Gamma\left(\nu/2\right)}\right]=0\\
    \Longrightarrow \quad  1+\log \frac{\nu}{2}-\psi \left(\frac{\nu}{2}\right)+\frac{1}{n}\sum_{t=1}^n\left(\langle \log u_{t} \rangle -\langle u_{t} \rangle \right)=0,
\end{gather}


\begin{thebibliography}{}
    
    \bibitem[\protect\citeauthoryear{Archambeau et al.}{Archambeau et al.}{2006}]{RPPCA-2006-Archambeau}				
		Archambeau, C., Delannay, N., \& Verleysen, M. (2006, June). Robust probabilistic projections. \emph{In Proceedings of the 23rd International conference on machine learning}, (pp. 33-40).
		
    
    \bibitem[\protect\citeauthoryear{Azman and Pathmanathan}{Azman and Pathmanathan}{2022}]{GLMLC-2022-Azman}	
		Azman, S., \& Pathmanathan, D. (2022). The GLM framework of the Lee–Carter model: a multi-country study. \emph{Journal of Applied Statistics}, 49(3), 752-763.
		
	\bibitem[\protect\citeauthoryear{Bishop and Nasrabadi}{Bishop and Nasrabadi}{2006}]{PRML-2006-Bishop}		
		Bishop, C. M., \& Nasrabadi, N. M. (2006).  \emph{Pattern recognition and machine learning} (Vol. 4, No. 4, p. 738), New York: springer.
    
    \bibitem[\protect\citeauthoryear{Brouhns et al.}{Brouhns et al.}{2002}]{PoiLC-2002-Brouhns}	Brouhns, N., Denuit, M., \& Vermunt, J. K. (2002). A Poisson log-bilinear regression approach to the construction of projected lifetables.	\emph{Insurance: Mathematics and economics}, 31(3), 373-393.

    \bibitem[\protect\citeauthoryear{Cairns et al.}{Cairns et al.}{2020}]{Covid-2020-Cairns}	
	Cairns, A. J., Blake, D. P., Kessler, A., \& Kessler, M. (2020). The impact of COVID-19 on future higher-age mortality. \emph{Available at SSRN 3606988}.

    \bibitem[\protect\citeauthoryear{Cairns et al.}{Cairns et al.}{2023}]{Covid-2023-Cairns}	
	Cairns, A. J., Blake, D. P., Kessler, A., Mathur, R \& Kessler, M. (2023). Covid-19 Mortality:
    The Proportionality Hypothesis. \emph{Working Paper}.
    
    \bibitem[\protect\citeauthoryear{Chan}{Chan}{2002}]{Stoch-2002-Chan}	
	Chan, W. S. (2002). Stochastic investment modelling: a multiple time-series approach.	\emph{British Actuarial Journal}, 8(3), 545-591.
    
    \bibitem[\protect\citeauthoryear{Chen et al.}{Chen et al.}{2009}]{RPPCA-2009-Chen}
		Chen, T., Martin, E., \& Montague, G. (2009). Robust probabilistic PCA with missing data and contribution analysis for outlier detection. \emph{Computational Statistics $\&$ Data Analysis}, 53(10), 3706-3716.

		
	\bibitem[\protect\citeauthoryear{Clarkson}{Clarkson}{1988}]{CPCA-1988-Clarkson}	
	Clarkson, D. B. (1988). Remark AS R71: A remark on algorithm AS 211. The FG diagonalization algorithm.	\emph{Journal of the Royal Statistical Society. Series C (Applied Statistics)}, 37(1), 147-151.

    \bibitem[\protect\citeauthoryear{Congressional Research Service}{Congressional Research Service}{2020}]{Wardeath-2020-CRS}	
	Congressional Research Service. (2020).	American War and Military Operations Casualties: Lists and Statistics. \emph{https://crsreports.congress.gov/product/pdf/RL/RL32492}.
	

    \bibitem[\protect\citeauthoryear{Deaton and Paxson}{Deaton and Paxson}{2004}]{Mor-2004-Deaton}	
		Deaton, A., \& Paxson, C. (2004). Mortality, income, and income inequality over time in Britain and the United States. \emph{Perspectives on the Economics of Aging}, 247-286.
	
	\bibitem[\protect\citeauthoryear{Delwarde et al.}{Delwarde et al.}{2007}]{NBLC-2007-Delwarde}	Delwarde, A., Denuit, M., \& Partrat, C. (2007). Negative binomial version of the Lee–Carter model for mortality forecasting.	\emph{Applied Stochastic Models in Business and Industry}, 23(5), 385-401.	
	
	\bibitem[\protect\citeauthoryear{Dempster et al.}{Dempster et al.}{1977}]{EM-1977-Dempster}
		Dempster, A. P., Laird, N. M., \& Rubin, D. B. (1977). Maximum likelihood from incomplete data via the EM algorithm. \emph{Journal of the Royal Statistical Society: Series B (Methodological)}, 39(1), 1-22.  
	
	\bibitem[\protect\citeauthoryear{Diao et al.}{Diao et al.}{2021}]{DSA-2021-Diao}	Diao, L., Meng, Y., \& Weng, C. (2021). A DSA Algorithm for Mortality Forecasting.	\emph{North American Actuarial Journal}, 25(3), 438-458.
	
	\bibitem[\protect\citeauthoryear{Goodman}{Goodman}{1979}]{IRWS-1979-Goodman}	
	Goodman, L. A. (1979). Simple models for the analysis of association in cross-classifications having ordered categories. \emph{Journal of the American Statistical Association}, 74(367), 537-552.
	
	\bibitem[\protect\citeauthoryear{Guo and Howard}{Guo and Howard}{2022}]{MyPPCA-2022-Guo}	Guo, Y., \& Bondell, H. (2022). On robust probabilistic principal component analysis using multivariate t-distributions. \emph{Communications in Statistics-Theory and Methods}, 1-19.
	
	\bibitem[\protect\citeauthoryear{Hastie et al.}{Hastie et al.}{2009}]{ESLII-2009-Hastie}	    Hastie, T., Tibshirani, R., \& Friedman, J. H. (2009). \emph{The elements statistical learning: data mining, inference, and prediction} (2nd ed.). New York: Springer.
	
	\bibitem[\protect\citeauthoryear{Huber}{Huber}{2004}]{Robust-2004-Huber}		
	Huber, P. J. (2004).  \emph{Robust statistics} (Vol. 523). John Wiley \& Sons.

    \bibitem[\protect\citeauthoryear{Human Mortality Database}{Human Mortality Database}{HMD}]{HMD-2023-HMD}	
	Human Mortality Database. (2023) Max Planck Institute for Demographic Research, University of California, Berkeley, and French Institute for Demographic Studies. Available at \emph{www.mortality.org}.
	
	\bibitem[\protect\citeauthoryear{Kibria and Joarder}{Kibria and Joarder}{2006}]{t-2006-Kibria}	
		Kibria, B. G., \& Joarder, A. H. (2006).  A short review of multivariate t-distribution. \emph{Journal of Statistical research}, 40(1), 59-72.
		
	\bibitem[\protect\citeauthoryear{Kleinow}{Kleinow}{2015}]{CAE-2015-Kleinow}	
	Kleinow, T. (2015). A common age effect model for the mortality of multiple populations.	\emph{Insurance: Mathematics and Economics}, 63, 147-152.	
		
	
	\bibitem[\protect\citeauthoryear{Künsch et al.}{Künsch et al.}{1989}]{OutlierPoi-1989-Kunsch}	
    Künsch, H. R., Stefanski, L. A., \& Carroll, R. J. (1989). Conditionally unbiased bounded-influence estimation in general regression models, with applications to generalized linear models.	\emph{Journal of the American Statistical Association}, 84(406), 460-466.
    
    \bibitem[\protect\citeauthoryear{Lange et al.}{Lange et al.}{1989}]{Robust-1989-Lange}	Lange, K. L., Little, R. J., \& Taylor, J. M. (1989). Robust statistical modelling using the t distribution.	\emph{Journal of the American Statistical Association}, 84(408), 881-896.
		
	\bibitem[\protect\citeauthoryear{Lee and Carter}{Lee and Carter}{1992}]{LC-1992}	
		Lee, R. D., \& Carter, L. R. (1992). modelling and forecasting US mortality. \emph{Journal of the American statistical association}, 87(419), 659-671.
		
		
	\bibitem[\protect\citeauthoryear{Li and Chan}{Li and Chan}{2005}]{OutlierLC-2005-Li}	
		Li, S. H., \& Chan, W. S. (2005). Outlier analysis and mortality forecasting: the United Kingdom and Scandinavian countries. \emph{Scandinavian Actuarial Journal}, 2005(3), 187-211.
		
	\bibitem[\protect\citeauthoryear{Li and Chan}{Li and Chan}{2007}]{OutlierLC-2007-Li}	
		Li, S. H., \& Chan, W. S. (2007). The Lee-Carter model for forecasting mortality, revisited. \emph{North American Actuarial Journal}, 11(1), 68-89.

    \bibitem[\protect\citeauthoryear{Li and Lee}{Li and Lee}{2005}]{multiLC-2005-Li}	
		Li, N., \& Lee, R. D. (2005). Coherent mortality forecasts for a group of populations: An extension of the Lee-Carter method. \emph{Demography}, 42(3), 575-594.	
	
	\bibitem[\protect\citeauthoryear{Lin}{Lin}{1972}]{t-1972-Lin} 
	Lin, P. E. (1972). Some characterizations of the multivariate t distribution. \emph{Journal of Multivariate Analysis}, 2(3), 339-344.
		
	\bibitem[\protect\citeauthoryear{Liu and Rubin}{Liu and Rubin}{1995}]{tEM-1995-Liu}
		Liu, C., \& Rubin, D. B. (1995). ML estimation of the t distribution using EM and its extensions, ECM and ECME. \emph{Statistica Sinica}, 19-39.
		
	\bibitem[\protect\citeauthoryear{Morgenthaler}{Morgenthaler}{1992}]{OutlierGLM-1992-Morgenthaler}	
	Morgenthaler, S. (1992). Least-absolute-deviations fits for generalized linear models. \emph{Biometrika}, 79(4), 747-754.

		
	\bibitem[\protect\citeauthoryear{Pedroza}{Pedroza}{2006}]{BayesLC-2006-Pedroza}	
		Pedroza, C. (2006). A Bayesian forecasting model: predicting US male mortality. \emph{Biostatistics}, 7(4), 530-550.
		
	\bibitem[\protect\citeauthoryear{Renshaw and Haberman}{Renshaw and Haberman}{2003}]{EnhLC-2003-Renshaw}	
		Renshaw, A. E., \& Haberman, S. (2003). Lee–Carter mortality forecasting with age-specific enhancement. \emph{Insurance: Mathematics and Economics}, 42(3), 575-594.
	
	\bibitem[\protect\citeauthoryear{Richman and Wüthrich}{Richman and Wüthrich}{2021}]{NNLC-2021-Richman}	
		Richman, R., \& Wüthrich, M. V. (2021). A neural network extension of the Lee–Carter model to multiple populations. \emph{Annals of Actuarial Science}, 15(2), 346-366.

    \bibitem[\protect\citeauthoryear{Smith}{Smith}{1947}]{WWII-1947-Smith}	
		Smith, M. (1947). Populational characteristics of American servicemen in World War II.  \emph{The Scientific Monthly}, 65(3), 246-252.
		
	\bibitem[\protect\citeauthoryear{Tipping and Bishop}{Tipping and Bishop}{1999}]{PPCA-1999-Tipping}				
		Tipping, M. E., \& Bishop, C. M. (1999). Probabilistic principal component analysis. \emph{Journal of the Royal Statistical Society: Series B (Statistical Methodology)}, 61(3), 611-622.

	\bibitem[\protect\citeauthoryear{Wang et al.}{Wang et al.}{2011}]{HTLC-2011-Wang}	
		Wang, C. W., Huang, H. C., \& Liu, I. C. (2011). A quantitative comparison of the Lee-Carter model under different types of non-Gaussian innovations. \emph{The Geneva Papers on Risk and Insurance-Issues and Practice}, 36(4), 675-696.
	 
\end{thebibliography}
\end{document}